\documentclass[twocolumn,aps,pre,superscriptaddress,showkeys,showpacs,floatfix]{revtex4-1} %% preprint,

\usepackage[T1]{fontenc}
\usepackage{amsmath}
\usepackage{graphicx}
\usepackage[font=small]{caption}
\usepackage[hypcap=true,font=footnotesize]{subcaption}

\usepackage[finalnew]{trackchanges}
\usepackage[colorlinks=true,hyperfootnotes=true,breaklinks=true]{hyperref}

\addeditor{R1}
\begin{document}
\def\arxiv{1}

\title{\change[R1]{Impact of the workers' loyalty on t}{T}he \add[R1]{working} group performance modeled by a bi-layer cellular automaton\remove[R1]{ with a hysteretic rule}}
\author{Krzysztof Malarz}
\homepage{http://home.agh.edu.pl/malarz/}
\email{malarz@agh.edu.pl}
\affiliation{\href{http://www.agh.edu.pl/}{AGH University of Science and Technology},
\href{http://www.pacs.agh.edu.pl/}{Faculty of Physics and Applied Computer Science},\\
al. Mickiewicza 30, 30-059 Krakow, Poland.}

\author{Agnieszka Kowalska-Stycze\'n}
\affiliation{\href{http://www.polsl.pl/}{Silesian University of Technology},
Faculty of Organization and Management,\\
ul. Roosevelta 26/28, 41-800 Zabrze, Poland.}

\author{Krzysztof Ku{\l}akowski}
\email{kulakowski@fis.agh.edu.pl}
\affiliation{\href{http://www.agh.edu.pl/}{AGH University of Science and Technology},
\href{http://www.pacs.agh.edu.pl/}{Faculty of Physics and Applied Computer Science},\\
al. Mickiewicza 30, 30-059 Krakow, Poland.}

\begin{abstract}
The problem `human and work' in a model working group is investigated by means of cellular automata technique.
Attitude of members of a group towards work is measured by an indicator of loyalty to the group (the number of agents who carry out their tasks), and lack of loyalty (the number of agents, who give their tasks to other agents).
Initially, all agents realize scheduled tasks one-by-one.
Agents with the number of scheduled tasks larger than a given threshold change their strategy to \change[R1]{`overloaded'}{unloyal} one and start avoiding completing tasks by passing them to their colleagues.
Optionally, in some conditions, we allow agents to return to \change[R1]{`underloaded'}{loyal} state; hence the rule is hysteretic.
Results are presented on an influence of {\em i)} the density of tasks, {\em ii)} the threshold number of tasks assigned to the agents' forcing him/her for strategy change on the system efficiency.
We show that a `black' scenario of the system stacking in a \add[R1]{`}jammed phase\add[R1]{'} (with all agents \change[R1]{being in overloaded state}{preferring unloyal strategy} and having plenty tasks scheduled for realization) may be avoided when return to \change[R1]{underloaded state}{loyalty} is allowed and either {\em i)} the number of agents chosen for task realization, or {\em ii)} the number of assigned tasks, or {\em iii)} the threshold value of assigned tasks, which force the agent to conversion from \change[R1]{underloaded}{loyal strategy} to \change[R1]{overloaded}{unloyal one}, or {\em iv)} the threshold value of tasks assigned to \change[R1]{overloaded}{unloyal} agent, which force him/her to task redistribution among his/her neighbors, are smartly chosen.
\end{abstract}

% 	 Structures and organization in complex systems


\date{\today}
\maketitle

%% ##########################################################
\section{Introduction}
%% ##########################################################

In modeling social phenomena, most important obstacle is the complexity of human mind. Individual decisions depend not only on the present status of the environment, but also on the whole history of a given person. For a descriptive theory of individual decisions see \cite{1}.  
A predictive theory needs causal relations, what undermines free will of human beings.
Yet, from the point of view of a modeler, the eternal discussion of the free will is somewhat vain: even in a fully deterministic world, to identify the boundary between `yes' and `no' in a complex multidimensional space of arguments, constructed in our memory, is a hopeless task. 
(For a recent and provocative formulation of the problem of free will see Ref.~\cite{2}.)
When collective effects are concerned, social modelers pin their hopes in the law of large numbers, where an individual can be reduced to a black box. How deep reduction is legitimated, depends heavily on the modeler's purpose; social libraries are filled up with self-defending proclamations.
For an outstandingly realistic approach we refer to writings of Bruce Edmonds \cite{3a}.
Also, a set of articles in Ref. \cite{3b} can be recommended.
Our position here is that it is worthwhile to try to imagine consequences of our memory for collective effects, even if no grounded scientific strategy justifies a concomitant set of assumptions.
Once again, to keep the task simple, at least computationally, cellular automata \add[R1]{(CA)} are invaluable.

In \change[R1]{cellular automata}{CA}, memory can be introduced directly by an enrichment of the rule, as it was done in the construction of a reversible automaton \cite{4}.
There, the cell state at time $t$ depends not only on the environment state at time $t-1$ but also on the state at time $t-2$.
Another approach is to make the rule dependent of the mean state variable, calculated over the whole system history \cite{5}.
Here we prefer to switch to a new rule, when a given condition is fulfilled by the cell state.
In this way the system refers to its memory in a dynamical way, which cannot be predicted before the simulation is performed; this characteristics of the problem is known as the computational irreducibility \cite{6}.
Further, the system can be switched back to the previous rule, if another condition is met.
\change[R1]{Up to our knowledge, this kind of hysteretic rule has not been applied yet, at least  in social simulations.}{This kind of hysteretic rule has been applied in man-machine systems, where different sets of rules have been switched on by different procedures} \cite{gor1,gor2}. \add[R1]{Here we apply it as to refer to the human ability to modify the cognitive context as to enable decisions} \cite{bruc}.
\change[R1]{However, i}{I}t remains in a general accordance with the psychological concept of scripts \cite{1}, which is activated once; further behavior of an individual is executed according to this script.

Many works that have appeared in the last decade, have shown the usefulness of CA  in the field of modeling various aspects of organization's management systems.
Human (group of people) and work (workflow) are key elements of the management system, thus, the analysis of these elements and their relationships is a major subject of research in this area.
Robbins {\em et al.} \cite{Robbins1997} {formulate the three `paths of research' in business management.
They are: workflow optimization (work), human group behavior (human), and human-things interaction (human and work).}

The study of group behavior using the \change[R1]{CA model}{model based on CA} is fairly common, because it is a tool tailored to the nature and dynamics of social processes.
Broad discussion of this subject was submitted by {\em inter alia} Hegselmann and Flache \cite{Hegselmann1998}.
What provide the universality of CA are emerging attempts to build models within the two other perspectives of research in the field of management.
For example, Hassan and Tucker \cite{Hassan2010} have demonstrated the use of CA to optimize locations of objects (facility layout problem).
Optimizing of rearranging---for example, machine in the production floor---leads to minimization of transport costs (time) and thus it is a typical task in the field of workflow optimization.
Thirumaran {\em et al.} \cite{Thirumaran2012} {proposed a simulation model which allows to support the analysis of changes in the business processes of customer service web portal.
In both approaches, the individual agents (\change[R1]{cellular automata}{the automaton} cells) represent the technical facilities and business rules and do not have a direct relationship with employees.

In this work we are interested in prospect of `human and work' research.
In this area, the main goal of researchers is to build CA models, which are tools allowing to explain the global behavior of the analyzed system based on local interactions within groups of employees, and between employees and broadly understood environment.
Agents in such models represent the people (employees) and the parameters (attributes).
The CA rules represent procedures, which shall be adopted on the basis of more or less simplified assumptions and/or socio-psychological theories and knowledge of the organization and management theory.

The theoretical model in this perspective was proposed by Bin and Zhang} \cite{Bin2006}.
{This model allows analyzing the impact of managerial decisions on the behavior of members of a group towards work---measured by specific indicator of loyalty to the group (in accordance with the general sociological message that a group loyal to each other is effective in achieving the objectives).
The authors assume, in accordance with the theory of `social exchange', that people are motivated by a desire for social status and respect as much as they are motivated by gains that are material and/or monetary.
They introduce, by splitting the group members into `economic beings' and `social beings', the possibility to analyze management policy, which consists of incentives of economic and/or social nature.
The direction of movement depends on the assumed policy and the type of agent (a social agent is attracted by social incentives, and economic agent by material ones).
The simulation  finishes  with a state of equilibrium, and the evaluation of the applied policy consists of  determining indicator loyalty value for a group for this state.

Another, more complex approach is presented in the work of Shengping and Bin} \cite{Shengping2007}.
{The proposed model takes into account the types of work performed by a group of employees.
The authors introduced the characteristics of the tasks on a scale from `hard' to `soft' work.
Hard work must be completed chiefly with technical ability and soft work must be done with social communication ability.
Employees were also categorized into groups on the scale of a continuum between `working hard' and `social'.
An interaction between neighboring agents consisting of `reconciling' the way they cooperate (on a scale hard/social) of each agent with ambient agents (on the principle of the adoption pattern of majority) has also been introduced.
Simulated at each step, the overall `state behavior' of the group, is therefore a function of many factors including his/her behavior in last time step, his/her neighbors' behavior, his/her properties, characteristic and state of the work.
Although the precise interpretation of such a complex indicator is difficult, in general, the level of `state behavior' reflects the (average) degree of positive attitude towards work and in a broad sense can be associated with the level of loyalty, which was proposed in Ref.}~\cite{Bin2006}.

Management policy analysis in the perspective of the work initiated in Ref.~\cite{Bin2006} {provides a model for the proposed work of Saravakos and Sirakoulis} \cite{Saravakos2014}.
The authors, in a simple implementation of CA \add[R1]{technique}, have proposed the characteristics of workers behavior in the seven-point scale from extremely negative to absolutely positive.
As the second dimension of agent features  indicator depicting each employee's insistence, his ability to remain `uninfluenced' by his coworkers was adopted.
Insistence takes values from one to five, where one denotes an employee who is highly influenced by his neighborhood and thus his behavior is determined by his coworkers; five is the total resistance to the influence of neighbors.
For intermediate sizes weights of impact were adopted.
Introduced behavioral rules allow to simulate (in each step) attitude change of each agent according to its attitude in the previous step, and the characteristics of its neighbors.
Additionally, in the model it was assumed that insistence is an adaptable trait and depends on the extent the employee is conformed to the organizational norms.
Organizational norms are determined by the company policy.
The organization's policy is represented by the `reward' coefficient (for positive behavior for agents with low insistence) and `punishment' coefficient (for negative behaviors and a large insistence).
Overall assessment of the behavior of group in a given environment is the average `loyalty factor'.
The main use of the model is to analyze the behavior of the tested group of workers (mainly by a loyalty factor) in the conditions of use of different combinations of reward and punishment levels. 

The general idea of the analysis of key factors and relationships in the term `human and work' and the ability to simulate specific managerial decision is also used in the proposed model in this article.
Loyalty is measured by the number of \change[R1]{underloaded}{loyal} agents who carry out their tasks, and the lack of loyalty is expressed by the number of \change[R1]{overloaded}{unloyal} agents, who give their tasks to other agents.

The latter activity is a channel of interaction between neighbors, what enables collective states.
The current strategy of a worker depends on the actual number of his/her awaiting tasks, what makes the changes dynamical and presumably complex.
Shifted tasks make the neighboring workers overburdened, what enhances the probability that they will also shift their tasks; this positive feedback allows to expect sharp transitions between collective states.
 
\add[R1]{A related problem in real life is the phenomenon of work stress induced burnout, which can lead to a selfish strategy to shifting duties to colleagues.  The problem is related to a number of professional groups, as nurses} \cite{nur} \add[R1]{and police officers} \cite{pol}. \add[R1]{The unloyal state takes the form of absenteeism and/or passivity. Basically, the transition to this state is irreversible; therefore the return to the loyal state, considered below, could be interpreted as the staff turnover.}

\add[R1]{At the group level, the opposition between the loyal and the unloyal phase is an example of a social dilemma, as defined in Ref.}~\cite{gro}\add[R1]{: a dichotomous choice of strategy, and the choice which is individually profitable, but makes worse when universally adopted. The concept of social dilemma is a generalization of the famous prisoner's dilemma, and it is used to discuss it in the frames of the game theory; for a simple introduction we recommend Ref.}~\cite{straf}. \add[R1]{Our assumption here is that the transition of an individual from the loyal to the unloyal phase is triggered by the large number of tasks. However, the unloyal behavior can be seen also as a reasonable strategy of self-preference, when the number of tasks exceeds some threshold. Thus, the frames of game theory enable yet another interpretation of  the transition from the unloyal back to the loyal state.}

In subsequent sections we define the cellular automaton (Sec.~\ref{sec-model}) and we present main results Sec.~\ref{sec-results} (including model verification Sec.~\ref{sec-verif}) obtained by means of computer simulations.
The Sec.~\ref{sec-disc} is devoted to the discussion of the results and conclusions.

%% ##########################################################
\section{\label{sec-model} Model}
%% ##########################################################

The model working group contains $L\times L$ agents occupying nodes of a square lattice.
Each agent may follow one of two strategies (loyal or unloyal to the group).
In the latter case an agent distributes his/her task among his/her nearest neighbors.
The loyal \remove[R1]{(underloaded)} agents do not bother their neighbors with their own task and during each time step they complete one of their tasks.
We start our simulations with group of \change[R1]{underloaded}{loyal} agents without any tasks assigned to them.
{Agent loyal to the group but} having more than some threshold number of scheduled tasks changes his/her strategy to {unloyal} one.
Unloyal \remove[R1]{(overloaded)} agents having assigned more than assumed threshold number of task shift part of their task to their adjacent neighbors.
These neighbors unconditionally take these additional tasks independently on the current number of tasks assigned to them.
The number of tasks awaiting for realization by a single agent cannot exceed the agent's maximal task capacity.
Incoming tasks are randomly distributed among different agents.
Finally, but optionally, we allow unloyal agents to convert to \change[R1]{underloaded state}{loyal ones}.
This step may be realized either as soon as {unloyal agent} shifts out all the tasks assigned to him/her or when \add[R1]{unloyal} agent \remove[R1]{in overloaded state} has not more than some threshold value of tasks.

To implement the model described above the \change[R1]{cellular automata}{CA} technique has been chosen.

A cellular automaton \cite{6,Wolfram,Ilacjhinski,Chopards} consists of a regular grid of cells, each in one of a finite number of states.
At each time step, a new generation of the cells' states is created, according to some fixed rule that determines the new state of each cell in terms of the current state of the cell and the states of the cells in its neighborhood.
Here, von Neumann neighborhood on $L\times L$ square bi-layer lattice with lateral periodic boundary conditions is assumed, {\em i.e.} the site $(i,j)$ have exactly $z=4$ neighbors at sites $(i\pm 1,j)$, $(i,j\pm 1)$.
The first lattice layer indicates agents strategy (\change[R1]{underloaded}{loyal} $X(i,j)=0$ or \change[R1]{overloaded state}{unloyal} $X(i,j)=1$).
The second layer carries information on current number of tasks $k(i,j)=0,\cdots, M$ assigned to $(i,j)$-th agent.
We start simulation with $L^2$ \change[R1]{underloaded}{loyal} agents awaiting for their first tasks, {\em i.e.}
\[
X(i,j)=0 \text{ and } k(i,j)=0 \text{ for } 1\le i,j\le L.
\]

The automata rule is as follow:
\begin{enumerate}
\item $K$ different sites $(i,j)$ are selected randomly.
The number of tasks assigned to agents occupying these sites is incremented by $Z$: 
\begin{subequations}
	\label{eq:step1}
	\begin{equation}
  	k_{t+1/6}(i,j)\equiv k_{t}(i,j)+Z,
	\end{equation}
	\begin{equation}
  	X_{t+1/6}(i,j)\equiv X_{t}(i,j).
	\end{equation}
\end{subequations}
\item Each \change[R1]{underloaded}{loyal} agent ($X_{t+1/6}(i,j)=0$) with more than $R$ tasks assigned to him/her ($k_{t+1/6}(i,j)>R$) changes his/her strategy to unloyal one:
\begin{subequations}
\label{eq:step2}
	\begin{equation}
  	X_{t+2/6}(i,j)\equiv 1,
	\end{equation}
	\begin{equation}
  	k_{t+2/6}(i,j)\equiv k_{t+1/6}(i,j).
	\end{equation}
\end{subequations}
\item Each \change[R1]{underloaded}{loyal} agent ($X_{t+2/6}(i,j)=0$) realizes one of his/her task:
\begin{subequations}
\label{eq:step3}
	\begin{equation}
  	k_{t+3/6}(i,j)\equiv\max\{k_{t+2/6}(i,j)-1,0\},
	\end{equation}
	\begin{equation}
  	X_{t+3/6}(i,j)\equiv X_{t+2/6}(i,j).
	\end{equation}
\end{subequations}
\item Each \add[R1]{unloyal} agent \remove[R1]{in overloaded state} ($X_{t+3/6}(i,j)=1$) with more than $T$ [and $(z-1)\le T\le R$] tasks assigned to him/her ($k_{t+3/6}(i,j)>T$) redistributes his/her own tasks among his/her ($z=4$) nearest neighbors
\begin{subequations}
\label{eq:step4}
	\begin{equation}
 	k_{t+4/6}(i,j)\equiv k_{t+3/6}(i,j)-z,
	\end{equation}
	\begin{equation}
 	X_{t+4/6}(i,j)\equiv X_{t+3/6}(i,j),
	\end{equation}
and
	\begin{equation}
 	k_{t+4/6}(i-1,j)\equiv k_{t+3/6}(i-1,j)+1,
	\end{equation}
	\begin{equation}
 	k_{t+4/6}(i+1,j)\equiv k_{t+3/6}(i+1,j)+1,
	\end{equation}
	\begin{equation}
 	k_{t+4/6}(i,j-1)\equiv k_{t+3/6}(i,j-1)+1,
	\end{equation}
	\begin{equation}
 	k_{t+4/6}(i,j+1)\equiv k_{t+3/6}(i,j+1)+1.
	\end{equation}
\end{subequations}
\item The number of tasks scheduled to a single agent cannot exceed maximal agent's capacity $M$:
\begin{subequations}
\label{eq:step5}
	\begin{equation}
 	k_{t+5/6}\equiv \min\{k_{t+4/6}(i,j),M\},
	\end{equation}
	\begin{equation}
 	X_{t+5/6}\equiv X_{t+4/6}(i,j).
	\end{equation}
\end{subequations}
\item The conversion of {unloyal agent} ($X_{t+5/6}(i,j)=1$) to {loyal member of the group} is possible
\begin{subequations}
\label{eq:step6}
	\begin{equation}
  	X_{t+6/6}(i,j)\equiv 0
	\end{equation}
	\begin{equation}
  	k_{t+6/6}(i,j)\equiv k_{t+5/6}(i,j)
	\end{equation}
\end{subequations}
	\begin{enumerate}
	\item[(A)] after shifting out all of his/her tasks ($k_{t+5/6}(i,j)=0$).
	\item[(B)] Alternatively, we allow for \change[R1]{overloaded}{unloyal} agents relaxation to \change[R1]{underloaded state}{loyal ones} when they have no more than $T$ tasks ($k_{t+5/6}(i,j)\le T$).
	\end{enumerate}
The steps \ref{eq:step6}(A) or \ref{eq:step6}(B) are realized only optionally.
We will show that realization of these steps is crucial for avoiding \add[R1]{`}black scenario\add[R1]{'} of {working group} total \add[R1]{`}jamming\add[R1]{'}.
\end{enumerate}
If steps \ref{eq:step6}(A) and \ref{eq:step6}(B) are omitted then simply
\[ X_{t+6/6}(i,j)\equiv X_{t+5/6}(i,j),\, k_{t+6/6}(i,j)\equiv k_{t+5/6}(i,j). \]
The steps described above are consecutively and synchronously applied to all sites.
The results for applying steps \ref{eq:step1}--\ref{eq:step5}, \ref{eq:step1}--\ref{eq:step5}+\ref{eq:step6}(A) and \ref{eq:step1}--\ref{eq:step5}+\ref{eq:step6}(B) are collected in Secs.~\ref{sec-irr}, \ref{sec-HR0} and \ref{sec-HRT}, respectively.

%% ##########################################################
\section{\label{sec-verif}Model verification}
%% ##########################################################
 
%% ----------------------------------------------------------
\begin{figure*}
  \if\arxiv0
\psfrag{t}{$t$}
  \fi
\begin{subfigure}[t]{0.99\textwidth}
  \includegraphics[width=0.95\textwidth]{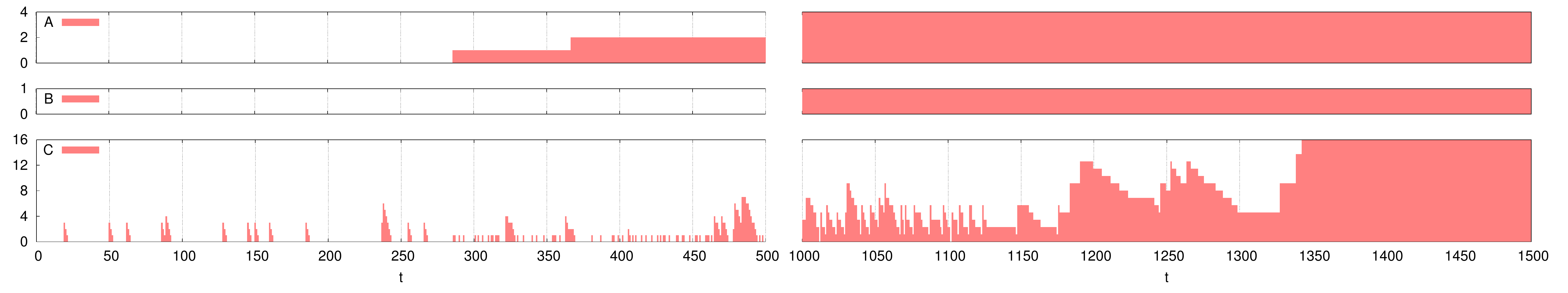}
  \caption{$K=5$} \label{single-a}
\end{subfigure}
\begin{subfigure}[t]{0.99\textwidth}
  \includegraphics[width=0.95\textwidth]{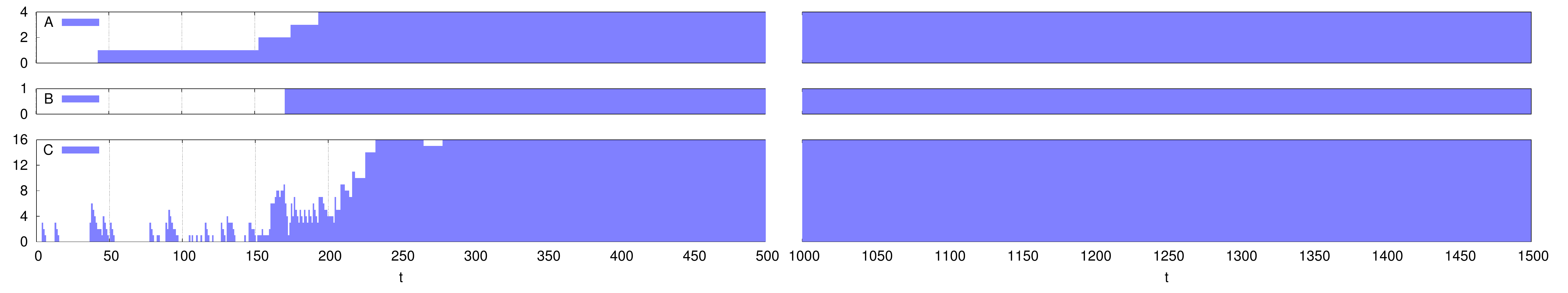}
  \caption{$K=8$} \label{single-b}
\end{subfigure}
\begin{subfigure}[t]{0.99\textwidth}
  \includegraphics[width=0.95\textwidth]{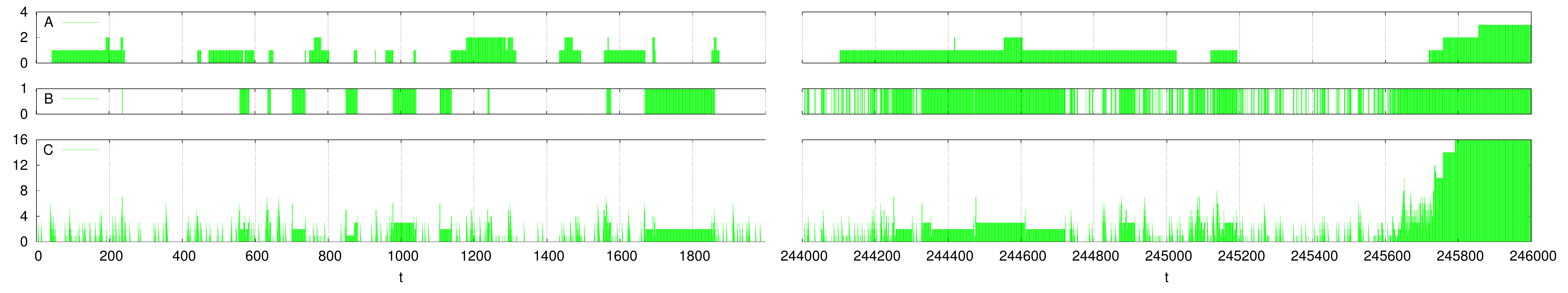}
  \caption{$K=5$} \label{single-c}
\end{subfigure}
\begin{subfigure}[t]{0.99\textwidth}
  \includegraphics[width=0.95\textwidth]{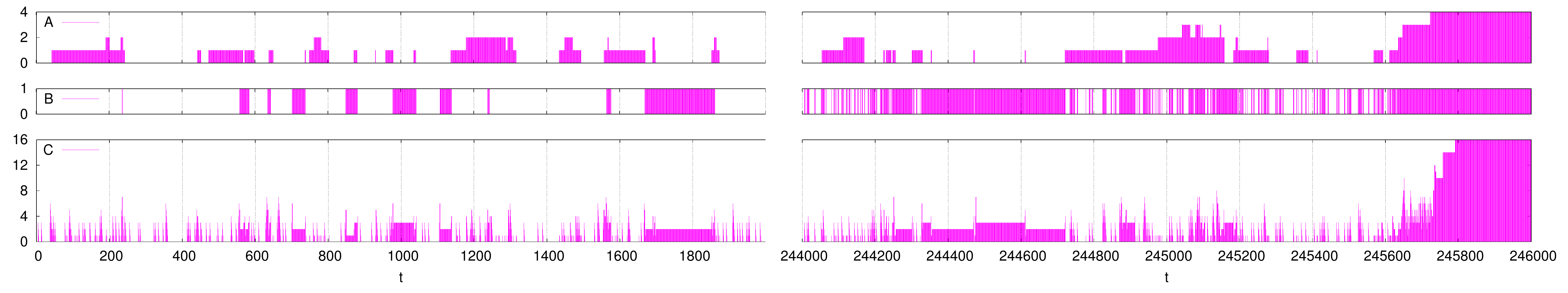}
  \caption{$K=8$} \label{single-d}
\end{subfigure}
\caption{\label{single}An example of the time evolution of (A) the number of the \change[R1]{overloaded}{unloyal} agents in agent's neighborhood $\ell$, (B) the agent state $X$ and (C) the number $k$ of tasks assigned for a single agent.
In simulations presented in panels (\subref{single-a}) and (\subref{single-b}) the back conversions of {unloyal agents} to \change[R1]{underloaded state}{loyal ones} were excluded.
In cases (\subref{single-c}) and (\subref{single-d}) the step \ref{eq:step6}(A) is realized.
$L=10$, $M=16$, $Z=4$, $R=8$, $T=3$.}
\end{figure*}
%% ----------------------------------------------------------

The model {verification} for computer models means the model evaluation or `process of reaching to sufficient confidence that the model is ready for use in particular case' \cite{Coyle}.
In order to {verify} \cite{3b} our model we inspect single agent temporal evolution of
\begin{enumerate}
 \item[(A)] the number $\ell$ of his/her \change[R1]{overloaded}{unloyal} nearest neighbors,
 \item[(B)] his/her strategy $X$,
 \item[(C)] and the number $k$ of tasks currently scheduled for her/him.
\end{enumerate}
These dependencies are shown in A, B, C panels of Fig.~\ref{single}.
Four sub-figures of Fig.~\ref{single} correspond to possibility of agents back conversion to \change[R1]{underloaded state}{loyalty} (\ref{single-c},\ref{single-d}) and reasonably (\ref{single-a}, \ref{single-c}, $K=5$) or enhanced (\ref{single-b}, \ref{single-d}, $K=8$) tasks delivering.
For these set of parameters ($L=10$, $M=16$, $R=8$, $Z=4$, $T=3$) the system becomes totally \add[R1]{`}jammed\add[R1]{'}: for long enough times of simulation all agents become \change[R1]{overloaded}{overburdened} and have to complete $M$ tasks.
The relaxation to \change[R1]{underloaded state}{loyalty} is realized according to the rule \ref{eq:step6}(A).
As expected the integer number of agents' {unloyal} nearest neighbors vary between 0 and $z$ ($\ell\in\{0,1,2,3,4=z\}$, panels A of Fig.~\ref{single}) and the number of tasks assigned to single agent cannot exceed \add[R1]{assumed} maximal number $M=16$ ($k\in\{0,1,\cdots,M-1,M\}$, panels C of Fig.~\ref{single}).
For irreversible transition to the unloyal \change[R1]{rule}{strategy} only single transition of state $X_t=0 \to X_{t+1}=1$ is expected as presented in panels B of Figs.~\ref{single-a} and \ref{single-b}.
On the contrary, for the rule \ref{eq:step6}(A) multiple transitions $X_t=0\to X_{t+1}=1$ and $X_t=1\to X_{t+1}=0$ may be observed [panels B of Figs.~\ref{single-c} and \ref{single-d}].

%% ##########################################################
\section{\label{sec-results}Results}
%% ##########################################################

%% ##########################################################
\subsection{\label{sec-irr}Irreversible transition to the unloyal\add[R1]{ty} \remove[R1]{agent rule}}
%% ##########################################################

\begin{figure*}
\includegraphics[width=0.8\textwidth]{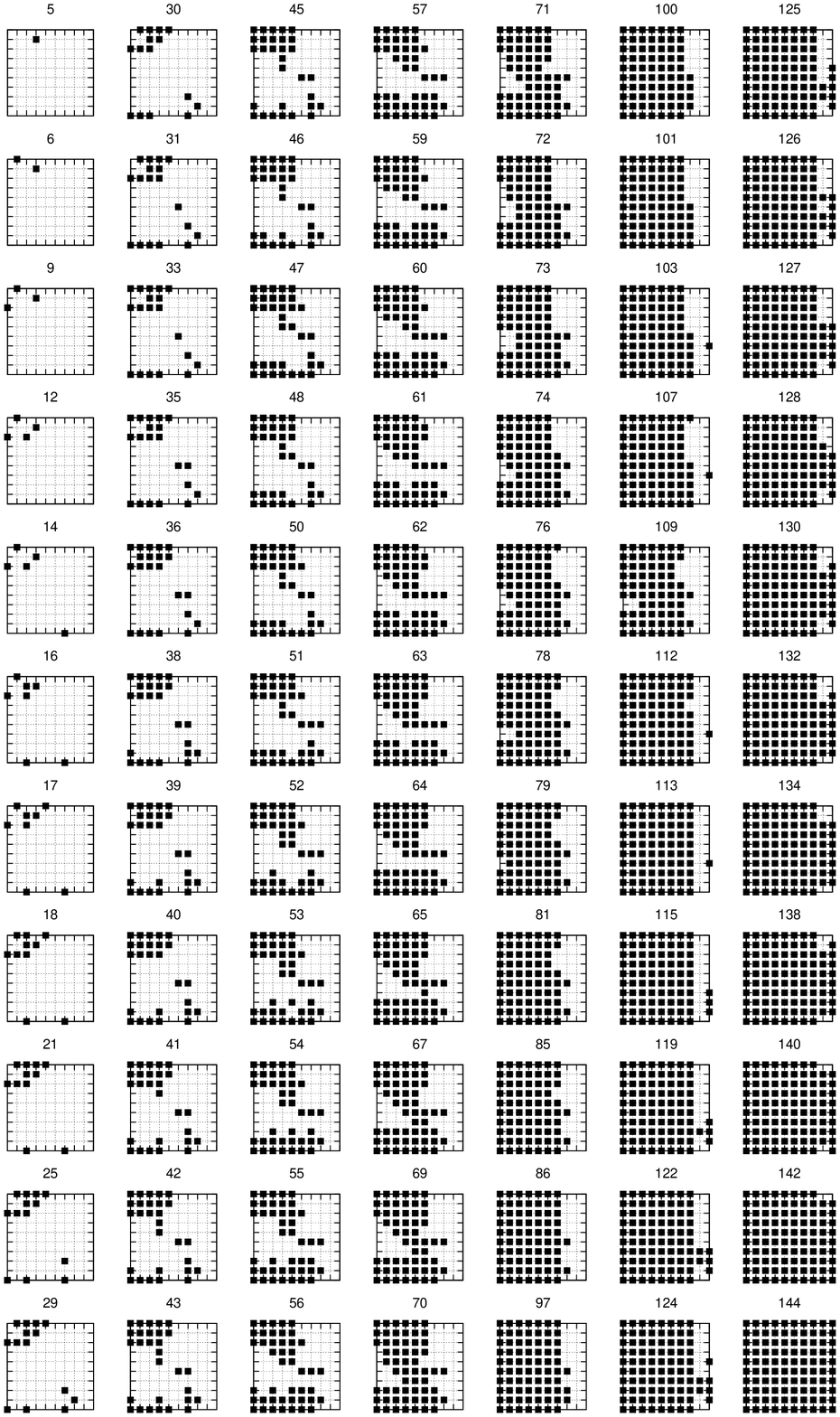}
\caption{\label{egoistsxy} The time evolution of spatial distribution of unloyal agents.
The back conversions of \change[R1]{overloaded}{unloyal} agents to \remove[R1]{underloaded} loyal \change[R1]{state}{ones} are excluded.
$L=10$, $M=16$, $Z=4$, $R=8$, $K=10$, $T=3$.}
\end{figure*}

In Fig.~\ref{egoistsxy} the time evolution of spatial distribution of \change[R1]{overloaded}{unloyal} agents is presented.
The back conversions of {unloyal agents} to \change[R1]{underloaded state}{loyal ones} are excluded, {\em i.e.} we do not realize the optional \ref{eq:step6}-th step from our algorithm presented in Sec. \ref{sec-model}.
The periodic boundary conditions are assumed.
After $t>20$ {unloyal agents} appear more likely at the border of existing \change[R1]{overloaded}{unloyal} agents' cluster.

%% ----------------------------------------------------------
%% ----------------------------------------------------------

\begin{figure}
  \if\arxiv1
     \includegraphics[width=0.95\columnwidth]{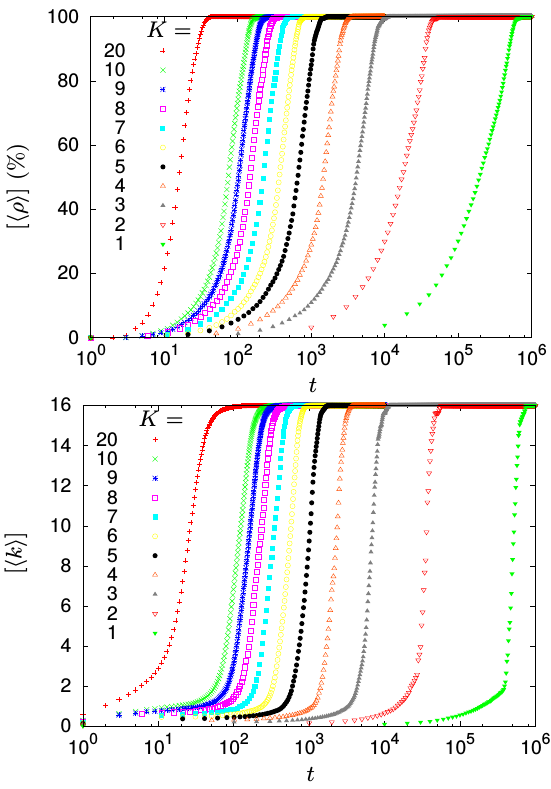}
  \else
\psfrag{K}{$K=$}
\psfrag{t}{$t$}
\psfrag{nk}{$[\langle k\rangle]$}
\psfrag{ne}[c]{$[\langle\rho\rangle]$ (\%)}
     \includegraphics[width=0.95\columnwidth]{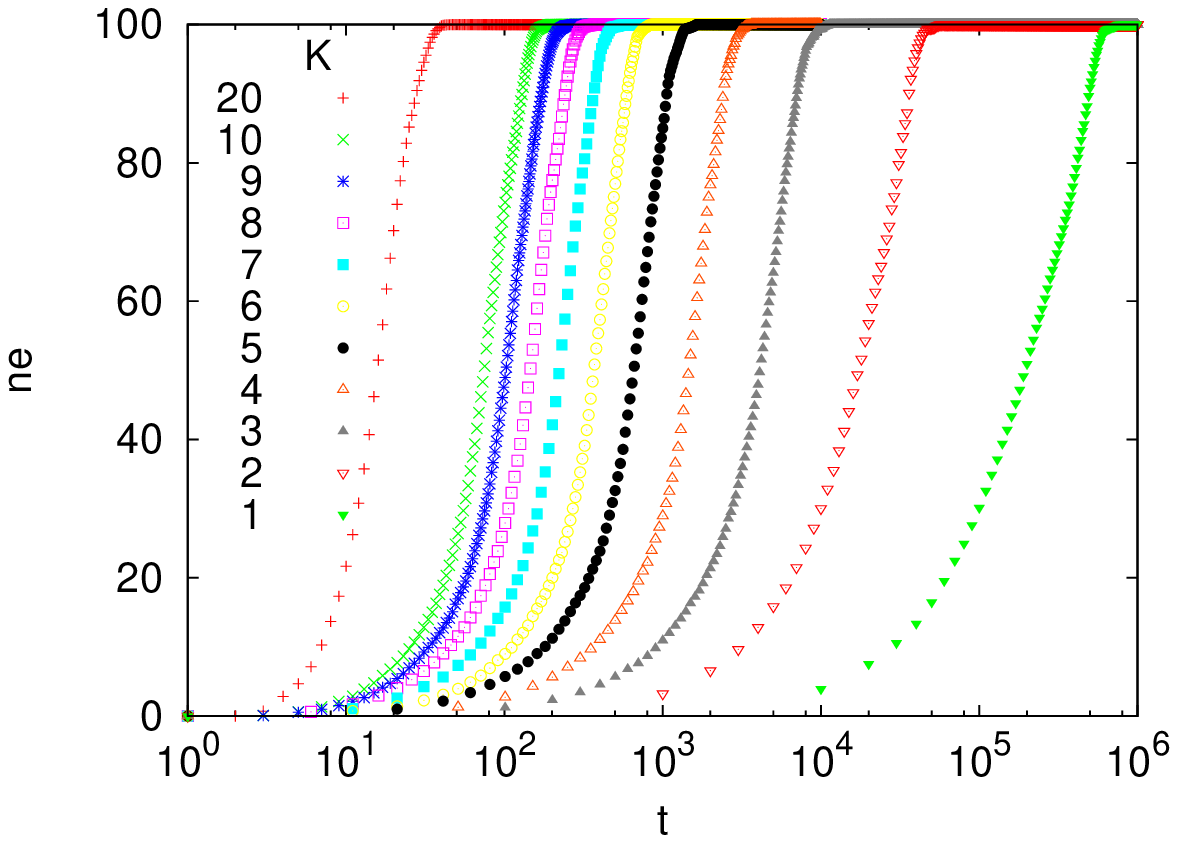}
     \includegraphics[width=0.95\columnwidth]{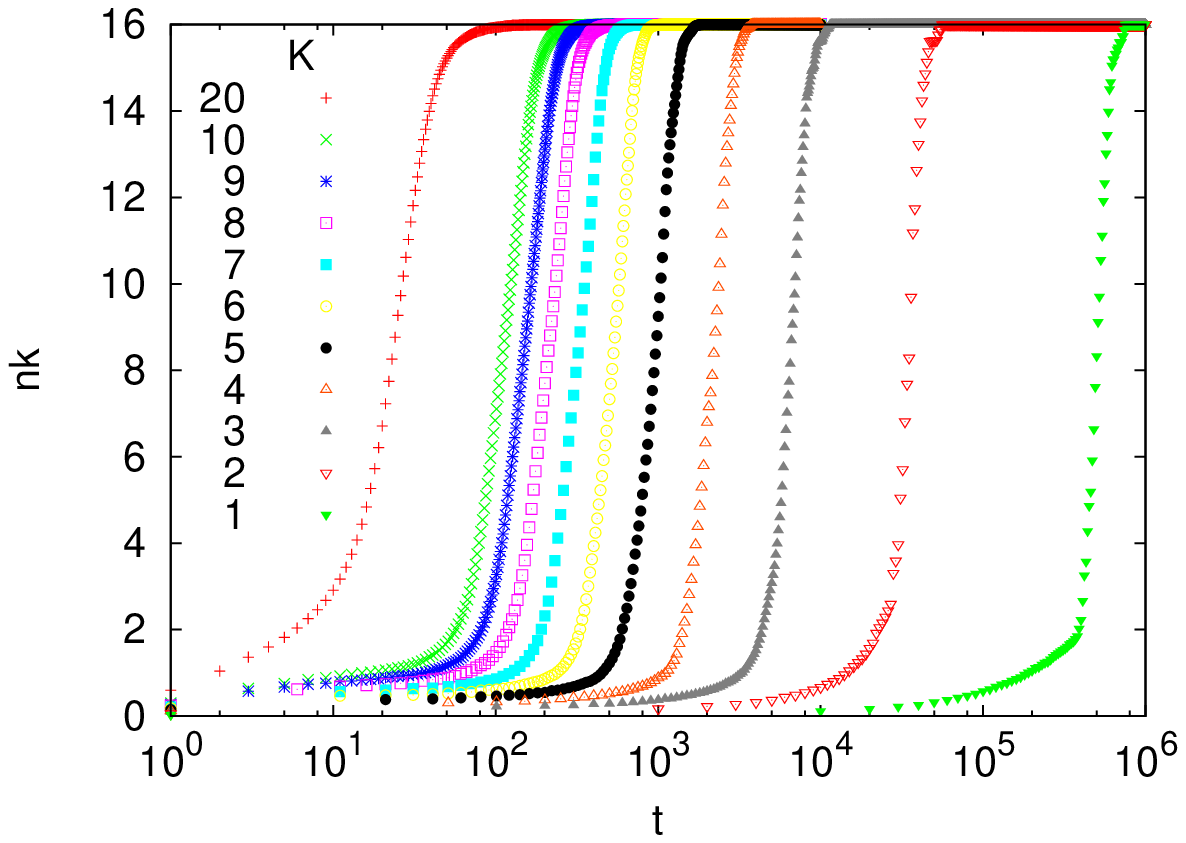}
  \fi
\caption{\label{fig-without-back-conversion} The time evolution of the average fraction of \change[R1]{overloaded}{unloyal} agents $[\langle\rho\rangle]$ (upper panel) and the average number of tasks per agent $[\langle k\rangle]$ (down panel).
The back conversions of \change[R1]{overloaded}{unloyal} agents to \change[R1]{underloaded states}{loyal ones} are excluded.
$L=10$, $M=16$, $Z=4$, $R=8$, $T=3$.
The results are averaged over $N=100$ simulations.}
\end{figure}

\begin{figure}
    \if\arxiv1
       \includegraphics[width=0.95\columnwidth]{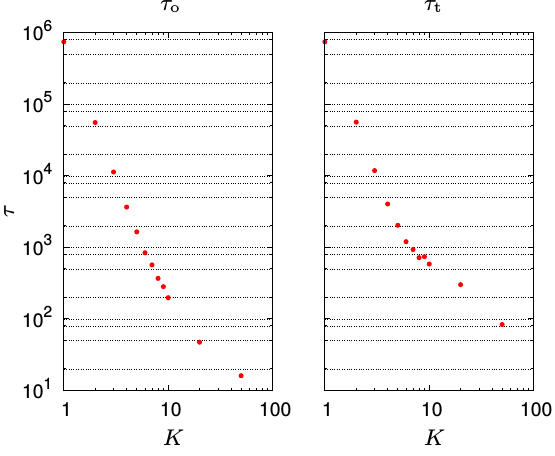}
    \else
\psfrag{K}{$K$}
\psfrag{tr}{$\tau$}
\psfrag{tt}{$\tau_{\text{t}}$}
\psfrag{te}{$\tau_{\text{o}}$}
     \includegraphics[width=0.95\columnwidth]{te_tt_vs_K}
    \fi
\caption{\label{fig-te-tt-vs-K} The time necessary to convert all agents to {unloyalty} ($\tau_{\text{o}}$) or to make them totally overburdened, \add[R1]{{\em i.e.} with $M$ tasks scheduled for realization,} ($\tau_{\text{t}}$) as a function of parameter $K$.
$L=10$, $M=16$, $Z=4$, $R=8$, $T=3$.
The results are averaged over $N=100$ independent simulations.}
\end{figure}

The time evolution of the average fraction of \change[R1]{overloaded}{unloyal} agents $[\langle\rho\rangle]$ (percent) and the average number of tasks per agent $[\langle k\rangle]$ are presented in Fig.~\ref{fig-without-back-conversion}.
Here, $\langle\cdots\rangle$ denotes a spatial average over all $L^2$ agents and $[\cdots ]$ stands for an average over $N$ different simulations.
Obviously, increasing number of sites $K$ where new tasks are delivered must lead to decreasing time ($\tau_\text{o}$) after which all agents become \change[R1]{overloaded}{unloyal} and to decreasing time ($\tau_\text{t}$) after which all agents have to complete maximally allowed number $M$ of tasks.
These times ($\tau_\text{o}$, $\tau_\text{t}$) dependencies on number $K$ of agents chosen for new $Z$ tasks realization are presented in Fig.~\ref{fig-te-tt-vs-K}. 

%% ----------------------------------------------------------
%% ----------------------------------------------------------
 
%% ----------------------------------------------------------
%% ----------------------------------------------------------

\add[R1]{Please note, that after time $t>\tau_{\text{t}}$ the newly incoming tasks are lost, as all both unloyal and overburdened agents ignore them.
This situation is caused by the assumed numerical technique (CA), in which single lattice cell may stay in finite number of states (here two, for the first automaton layer, and $M$ for the second one).
This however, does not influence our results qualitatively, as increasing of maximal agents capacity $M$ results only in a delay in reaching overburden state of all agents (see Fig.~}\ref{fig-tt-vs-M-a}\add[R1]{); the time of conversion all agents to unloyal group members ($\tau_{\text{t}}$) remains unchanged.
Time $\tau_{\text{t}}$ grows with $M$ roughly linearly (see Fig.~}\ref{fig-tt-vs-M-b}\add[R1]{).}

%% ----------------------------------------------------------
\begin{figure}
  \if\arxiv1
\begin{subfigure}[t]{0.99\columnwidth}
     \includegraphics[width=0.95\columnwidth]{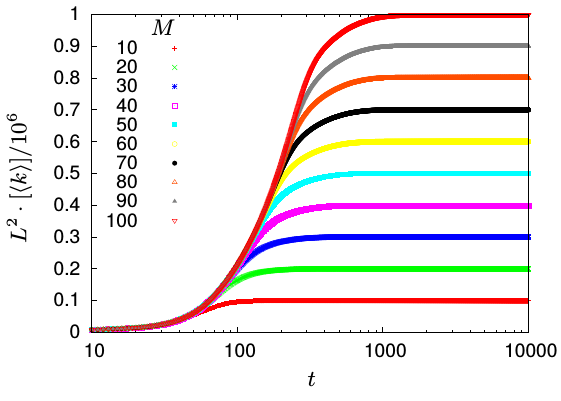}
     \caption{\label{fig-tt-vs-M-a}}
\end{subfigure}
\begin{subfigure}[t]{0.99\columnwidth}
     \includegraphics[width=0.95\columnwidth]{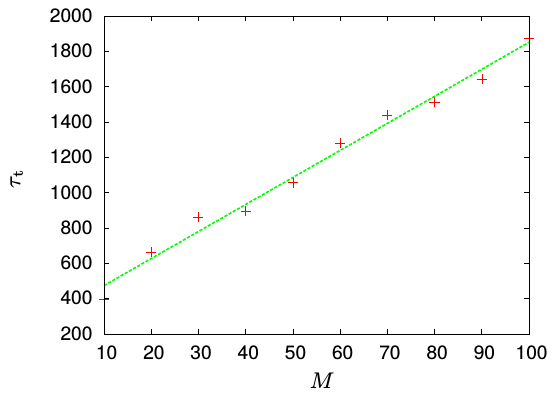}
     \caption{\label{fig-tt-vs-M-b}}
\end{subfigure}
  \else
\psfrag{M}{$M$}
\psfrag{t}{$t$}
\psfrag{y}[c]{$L^2\cdot [\langle k\rangle]/10^6$}
\psfrag{tau}{$\tau_{\text{t}}$}
\begin{subfigure}[t]{0.99\columnwidth}
     \includegraphics[width=0.95\columnwidth]{y_vs_t_on_M}
     \caption{\label{fig-tt-vs-M-a}}
\end{subfigure}
\begin{subfigure}[t]{0.99\columnwidth}
     \includegraphics[width=0.95\columnwidth]{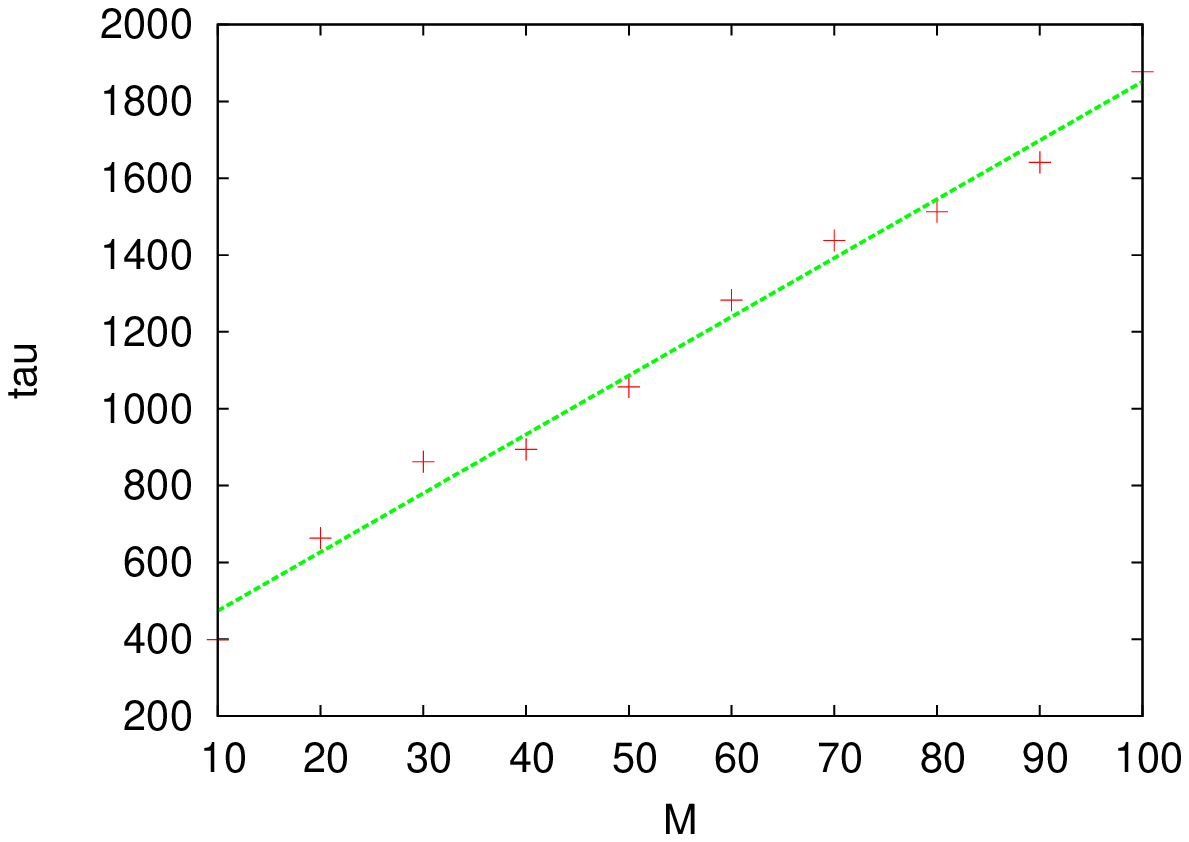}
     \caption{\label{fig-tt-vs-M-b}}
\end{subfigure}
  \fi
\caption{\label{fig-tt-vs-M} \add[R1]{(Color online)
(a) Time evolution of the total number $L^2\cdot [\langle k\rangle]$ of tasks scheduled for realization for various values of maximal agents' capacity $M$.
(b) The time necessary to make all agents totally overburdened ($\tau_{\text{t}}$) as a function of parameter $M$.
$L=10$, $Z=4$, $R=4$, $K=8$, $T=3$.
The results are averaged over $N=100$ independent simulations.}}
\end{figure}
%% ----------------------------------------------------------
 
%% ----------------------------------------------------------
\begin{figure}
  \if\arxiv1
     \includegraphics[width=0.95\columnwidth]{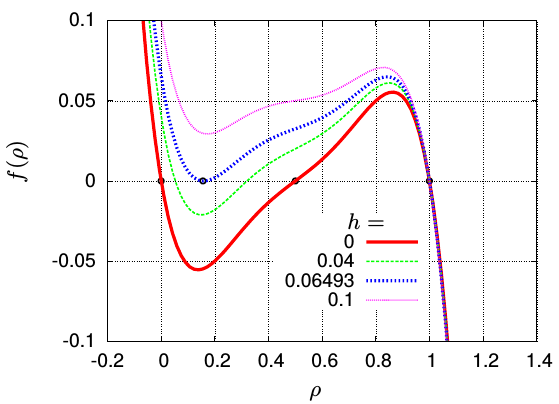}
  \else
\psfrag{f}{$f(\rho)$}
\psfrag{rho}{$\rho$}
\psfrag{h}{$h=$}
     \includegraphics[width=0.95\columnwidth]{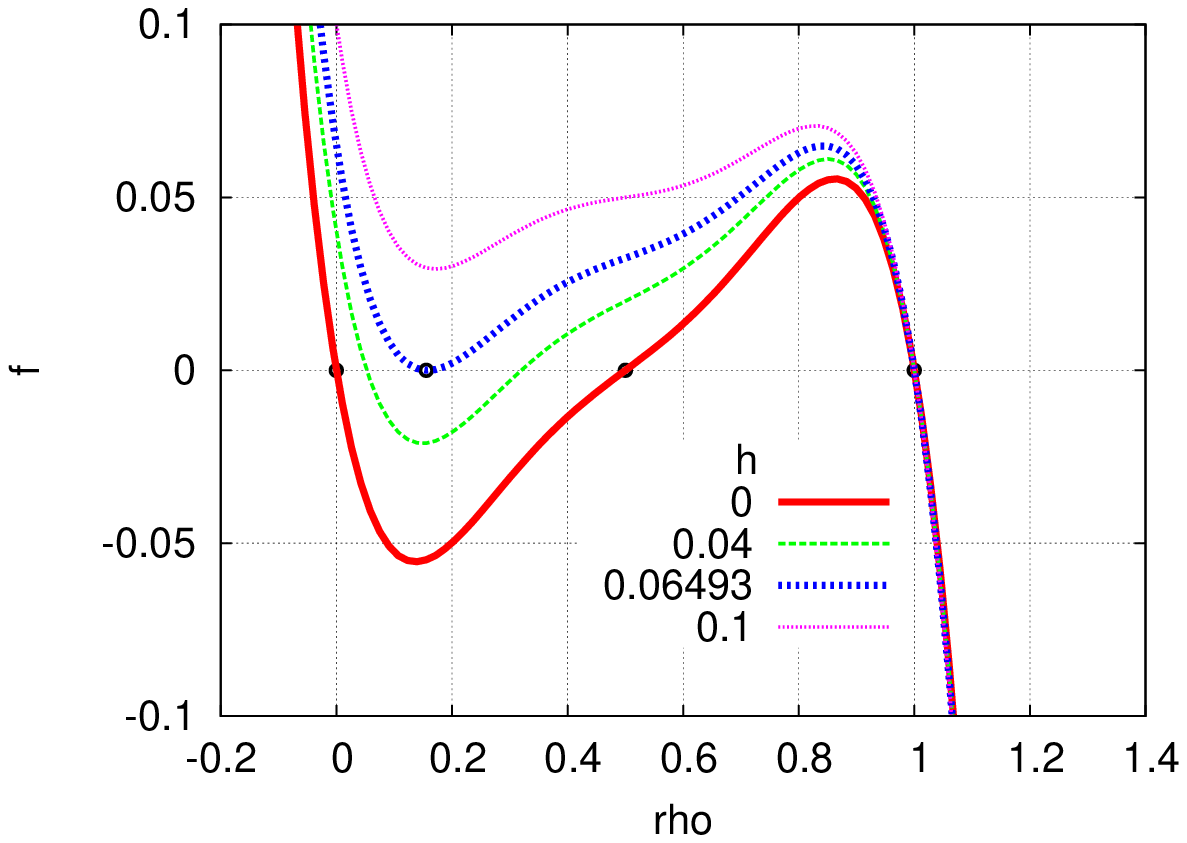}
  \fi
\caption{\label{fig-bif} \add[R1]{The r.h.s. of Eq.} \eqref{eq-mfa} \add[R1]{for various values of field $h$.
The saddle-node bifurcation appears when $f(\rho)=0$ and $f'(\rho)=0$, what defines the bifurcation field $h_b$.}}
\end{figure}
%% ----------------------------------------------------------

\add[R1]{This transition can be verified by a comparison with a simplified picture, obtained with a mean field model of a square lattice of agents.
The equation is composed in the spirit of Refs.} \cite{mfa,bag}, \add[R1]{where the set of states is reduced to unloyal (with probability $\rho$) and loyal (with probability $1-\rho$);}
\begin{equation}
\label{eq-mfa}
\begin{split}
\dot \rho = f(\rho)=\\
\sum_{i=0}^2 {4 \choose i}\big[ (1-\rho)^{1+i}\rho^{4-i} -\rho^{1+i}(1-\rho)^{4-i}\big]
%%(1-\rho)\rho^4 + {4 \choose 1}(1-\rho)^2\rho^3
%% -\rho(1-\rho)^4 -{4 \choose 1}\rho^2(1-\rho)^3\\
%% +{4 \choose 2}x^2(1-x)^3 -{4 \choose 2}x^3(1-x)^2
 +(1-\rho) h,
\end{split}
\end{equation}
\add[R1]{where the terms on the r.h.s. of Eq.} \eqref{eq-mfa} \add[R1]{for $i=0$ are related to a creation/anihilation of unloyal agent in the neighborhood of four unloyal/loyal ones, the terms for $i=1$ describe a creation/anihilation of unloyal agent in the neighborhood of three unloyal/loyal ones, etc.
The last term (field $h>0$) is a creation of unloyal agent because of an external flow of tasks} \cite{mfa}.

\add[R1]{For $h=0$ we get three fixed points: $\rho^*_0 = 1$, $\rho^*_1=0$ (both stable) and $\rho^*_2=\frac{1}{2}$ (unstable).
When $h$ increases to $h_b\approx 0.06493966\cdots$, the roots $\rho^*_1$ and $\rho^*_2$ merge at the saddle-node bifurcation (see Fig.}~\ref{fig-bif}\add[R1]{) and the unloyal state $\rho^*_0 = 1$ remains as the only solution.
%% h_b = 0.06493966202
The same situation appears for the chain of agents and for the Bethe lattice with three neighbors, but with $h_b=\frac{1}{8}$.
Decrease of $h_b$ with increasing lattice coordination number seems to be natural, as for larger number of neighbors the transition to unloyal state should appear earlier.
Although much details are lost in this description, the basic result---the destabilization of the loyal state with an increasing input of tasks---is reproduced.}

%% ##########################################################
\subsection{Hysteretic rules}
%% ##########################################################

When back conversion of \change[R1]{overloaded}{unloyal agents} to \change[R1]{underloaded agents}{loyal ones} is excluded it is just a matter of time when the system will be totally overburden.
Without agents relaxation to \change[R1]{underloaded state}{loyalty} ($X=0$) the system inescapably tends to the situation where all agents are \change[R1]{overloaded}{unloyal} and always have to complete yet $M$ tasks.

In order to avoid this `black scenario' {unloyal agents} must have a chance for relaxation to \change[R1]{underloaded}{the} state \add[R1]{of loyalty}.
However, even when back conversion is allowed for some set of parameters the system \add[R1]{`}jamming\add[R1]{'} may occur.

When \change[R1]{overloaded}{unloyal} agents are allowed to relax to the \change[R1]{underloaded}{loyalty} \remove[R1]{state} the ultimate system fate depends on assumed values of parameters ($K$, $T$, $R$ and $Z$).
The phase space of parameters $(K,T,R,Z)$ may be divided into two regions for which system
\begin{itemize}
\item either tends to the `jamming' state 
	\begin{equation}
	\lim_{t\to\infty}[\langle k\rangle]=M, \qquad \lim_{t\to\infty}[\langle\rho\rangle]=1,
	\end{equation}
\item or the average number of task awaiting for realization and the average fraction of \change[R1]{overloaded}{unloyal} agents do not reach their maximal values 
	\begin{equation}
	\lim_{t\to\infty}[\langle k\rangle]<M, \qquad \lim_{t\to\infty}[\langle\rho\rangle]<1.
	\end{equation}
\end{itemize}
In terms of statistical physics we can talk about `organizational' phase transition between `jamming' and `making-it' phases.
The efficiency of tasks realization in `making-it' phase depends quantitatively on applied relaxation rule [\ref{eq:step6}(A) or \ref{eq:step6}(B)].
The results of simulations for these two rules will be presented in subsequent two subsections \ref{sec-HR0} and \ref{sec-HRT}, respectively.

%% ##########################################################
\subsubsection{\label{sec-HR0} Returning to \change[R1]{underloaded state}{loyalty} after completing all tasks ($k=0$)}
%% ##########################################################

\begin{figure*}
  \if\arxiv1
\begin{subfigure}[t]{0.99\textwidth}
  \includegraphics[width=0.99\textwidth]{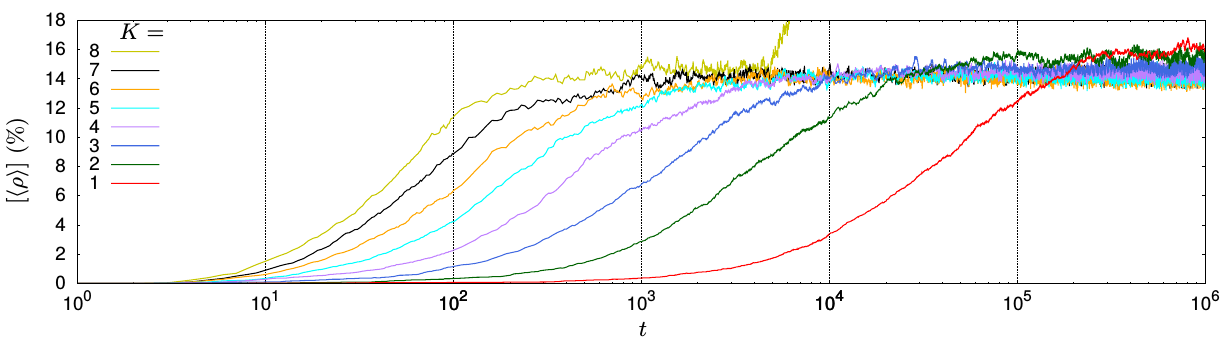}
  \caption{$M=16$, $Z=4$, $R=8$, $T=3$ and various values of $K$}\label{fig-egoists-K}
\end{subfigure}
\begin{subfigure}[t]{0.99\textwidth}
  \includegraphics[width=0.99\textwidth]{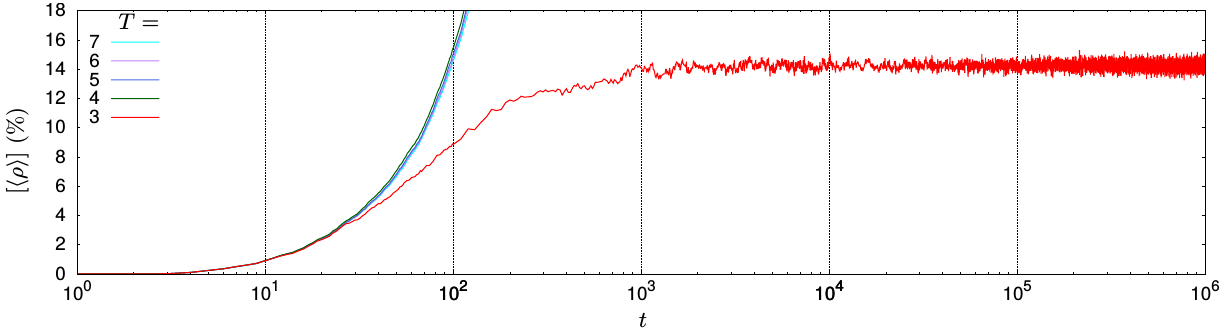}
  \caption{$M=16$, $Z=4$, $R=8$, $K=7$ and various values of $T$}\label{fig-egoists-T}
\end{subfigure}
\begin{subfigure}[t]{0.99\textwidth}
  \includegraphics[width=0.99\textwidth]{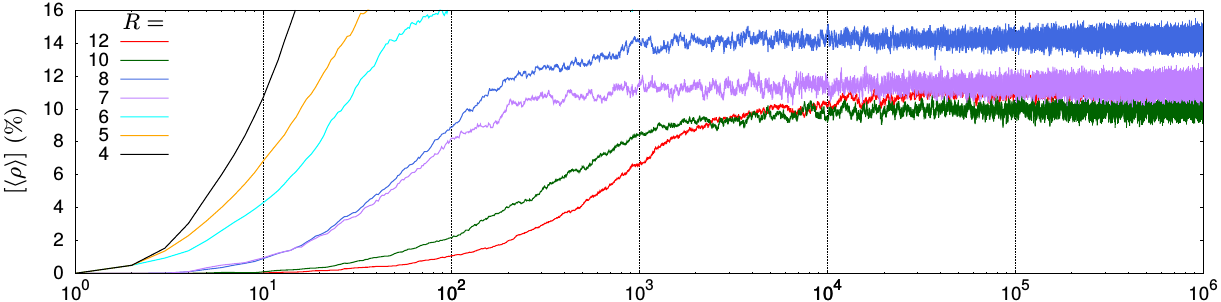}
  \caption{$M=16$, $Z=4$, $T=3$, $K=7$ and various values of $R$}\label{fig-egoists-R}
\end{subfigure}
\begin{subfigure}[t]{0.99\textwidth}
  \includegraphics[width=0.99\textwidth]{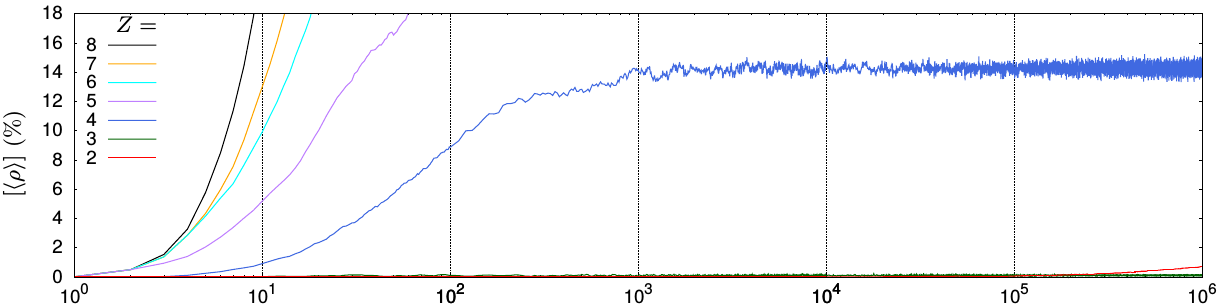}
  \caption{$M=16$, $R=8$, $T=3$, $K=7$ and various values of $Z$}\label{fig-egoists-Z}
\end{subfigure}
  \else
\psfrag{K=}{$K=$}
\psfrag{T=}{$T=$}
\psfrag{R=}{$R=$}
\psfrag{Z=}{$Z=$}
\psfrag{t}{$t$}
\psfrag{ne}[c]{$[\langle\rho\rangle]$ (\%)} 
\begin{subfigure}[t]{0.99\textwidth}
  \includegraphics[width=0.99\textwidth]{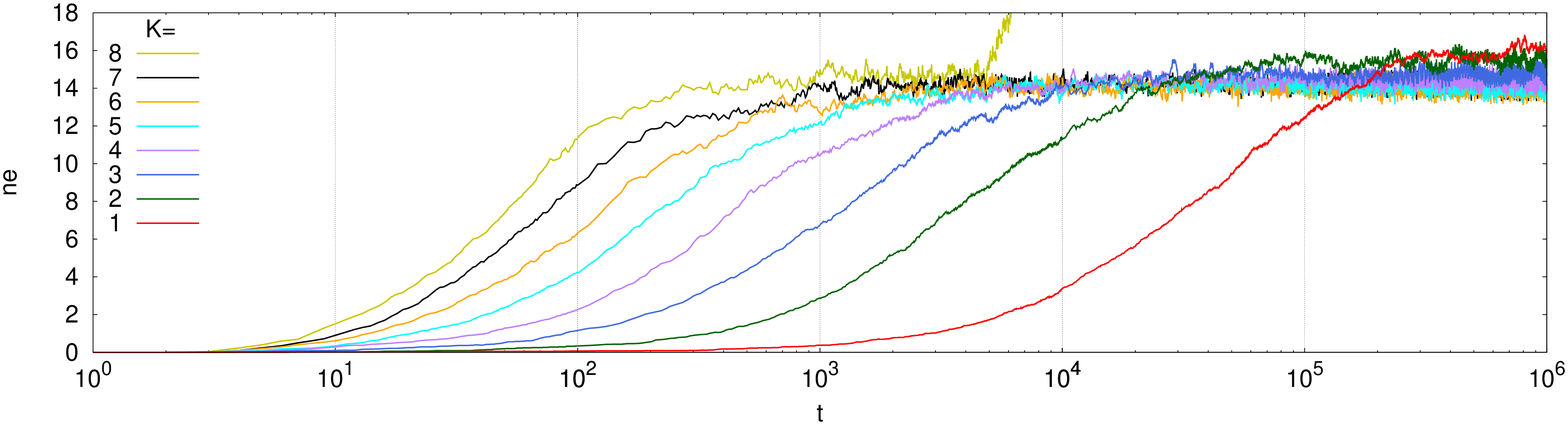}
  \caption{$M=16$, $Z=4$, $R=8$, $T=3$ and various values of $K$}\label{fig-egoists-K}
\end{subfigure}
\begin{subfigure}[t]{0.99\textwidth}
  \includegraphics[width=0.99\textwidth]{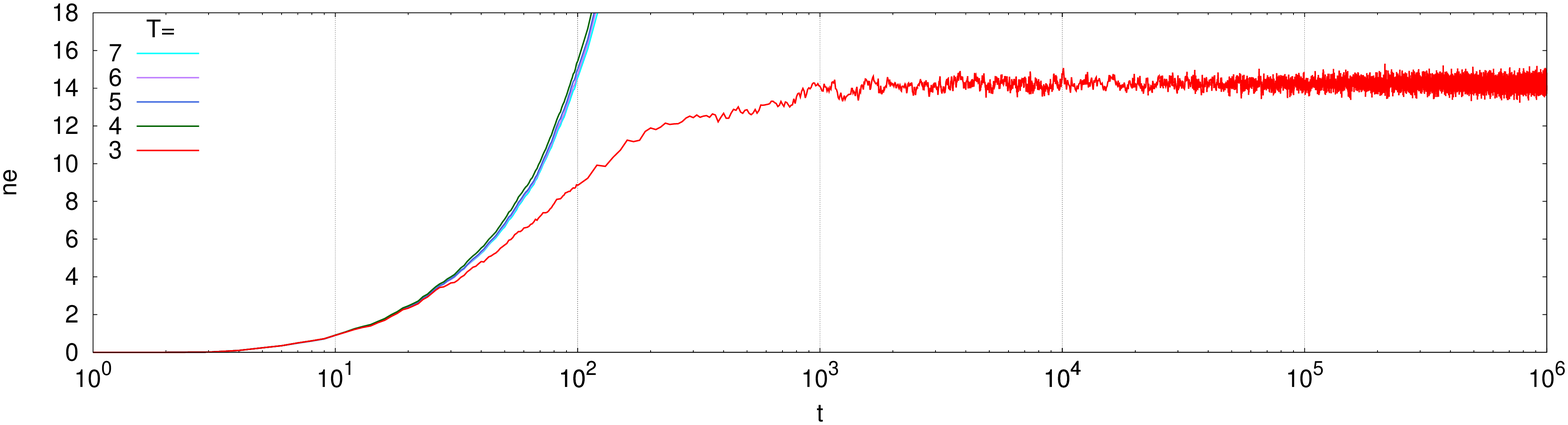}
  \caption{$M=16$, $Z=4$, $R=8$, $K=7$ and various values of $T$}\label{fig-egoists-T}
\end{subfigure}
\begin{subfigure}[t]{0.99\textwidth}
  \includegraphics[width=0.99\textwidth]{HR0_L10_M16_Z4_K7_T3_R_negoists}
  \caption{$M=16$, $Z=4$, $T=3$, $K=7$ and various values of $R$}\label{fig-egoists-R}
\end{subfigure}
\begin{subfigure}[t]{0.99\textwidth}
  \includegraphics[width=0.99\textwidth]{HR0_L10_M16_R8_T3_K7_Z_negoists}
  \caption{$M=16$, $R=8$, $T=3$, $K=7$ and various values of $Z$}\label{fig-egoists-Z}
\end{subfigure}
  \fi
\caption{\label{fig-egoists} (Color online). Time evolution of the average fraction $[\langle\rho\rangle]$ of \change[R1]{overloaded}{unloyal} agents obtained for \ref{eq:step6}(A) rule.
$L=10$.
The results are averaged over $N=100$ simulations.}
\end{figure*}

\begin{figure*}

  \if\arxiv1
\begin{subfigure}[t]{0.99\textwidth}
  {\includegraphics[width=0.99\textwidth]{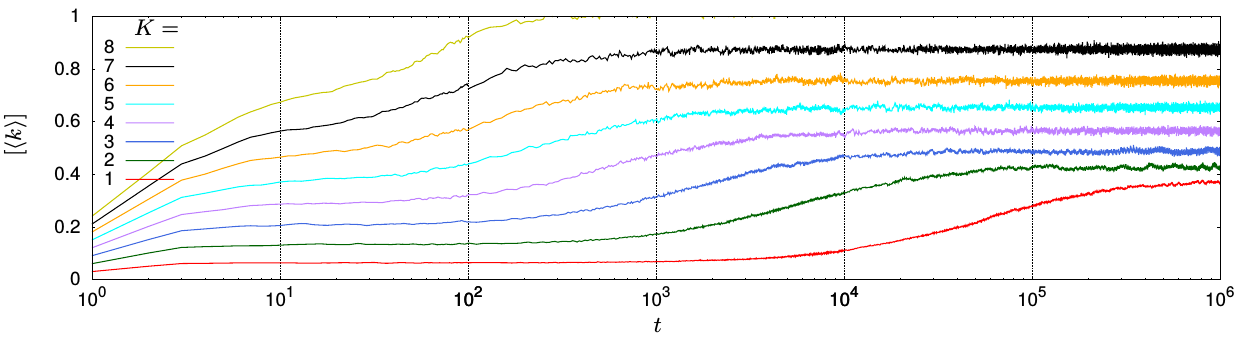}}
     \caption{\label{fig-tasks-K} $M=16$, $Z=4$, $R=8$, $T=3$ and various values of $K$}
\end{subfigure}
\begin{subfigure}[t]{0.99\textwidth}
  {\includegraphics[width=0.99\textwidth]{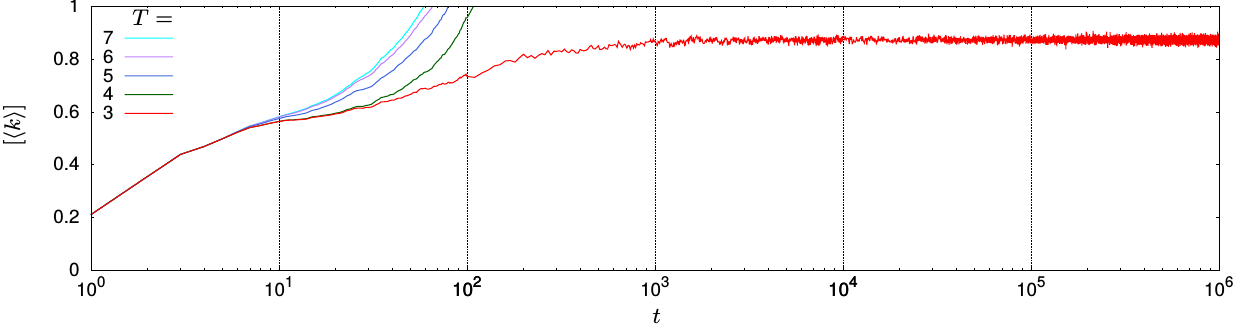}}
     \caption{\label{fig-tasks-T} $M=16$, $Z=4$, $R=8$, $K=7$ and various values of $T$}
\end{subfigure}
\begin{subfigure}[t]{0.99\textwidth}
  {\includegraphics[width=0.99\textwidth]{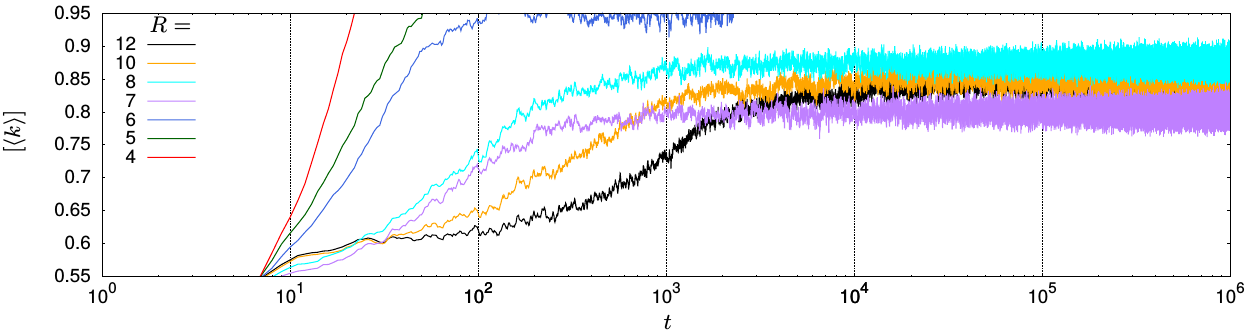}}
     \caption{\label{fig-tasks-R} $M=16$, $Z=4$, $T=3$, $K=7$ and various values of $R$}
\end{subfigure}
\begin{subfigure}[t]{0.99\textwidth}
  {\includegraphics[width=0.99\textwidth]{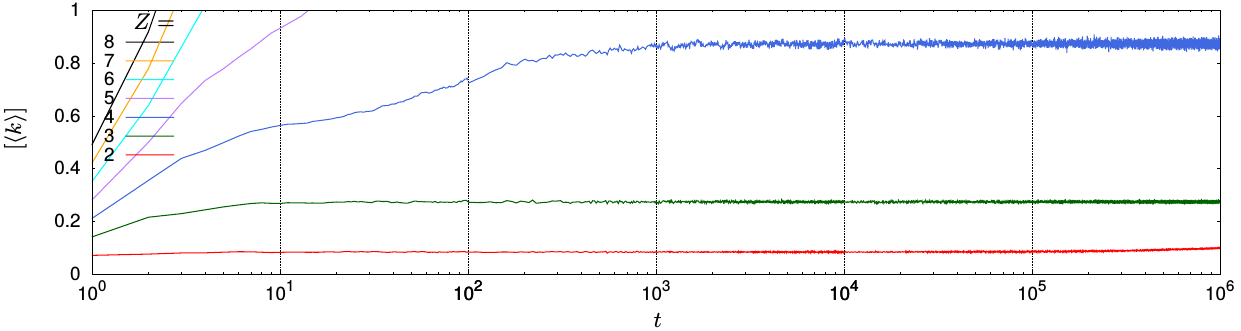}}
     \caption{\label{fig-tasks-Z} $M=16$, $R=8$, $T=3$, $K=7$ and various values of $Z$}
\end{subfigure}
  \else
\psfrag{K=}{$K=$}
\psfrag{T=}{$T=$}
\psfrag{R=}{$R=$}
\psfrag{Z=}{$Z=$}
\psfrag{t}{$t$}
\psfrag{nk}{$[\langle k\rangle]$}
\begin{subfigure}[t]{0.99\textwidth}
     \includegraphics[width=0.99\textwidth]{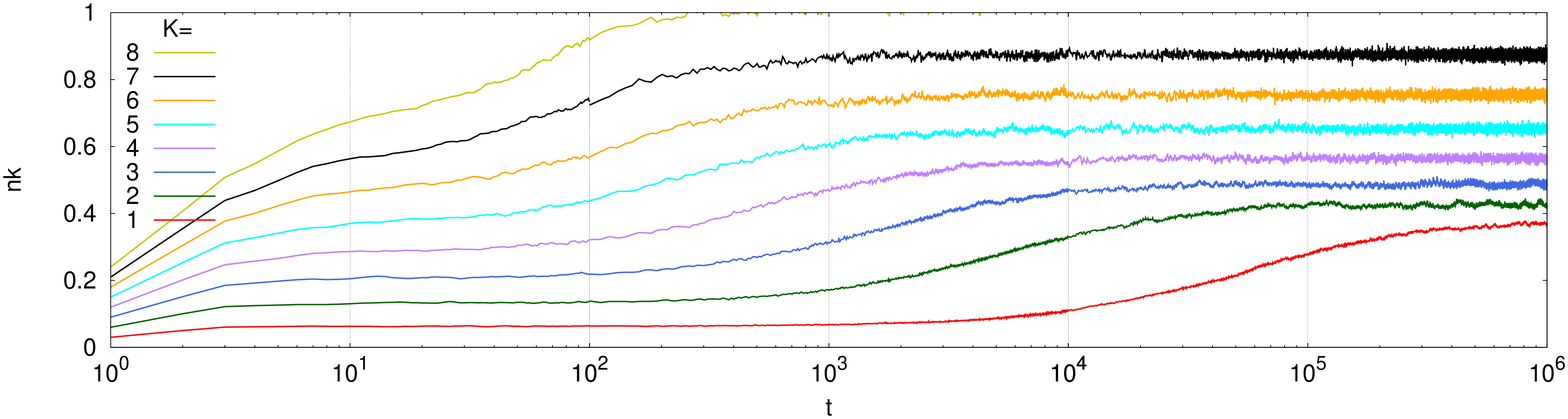}
     \caption{\label{fig-tasks-K} $M=16$, $Z=4$, $R=8$, $T=3$ and various values of $K$}
\end{subfigure}
\begin{subfigure}[t]{0.99\textwidth}
     \includegraphics[width=0.99\textwidth]{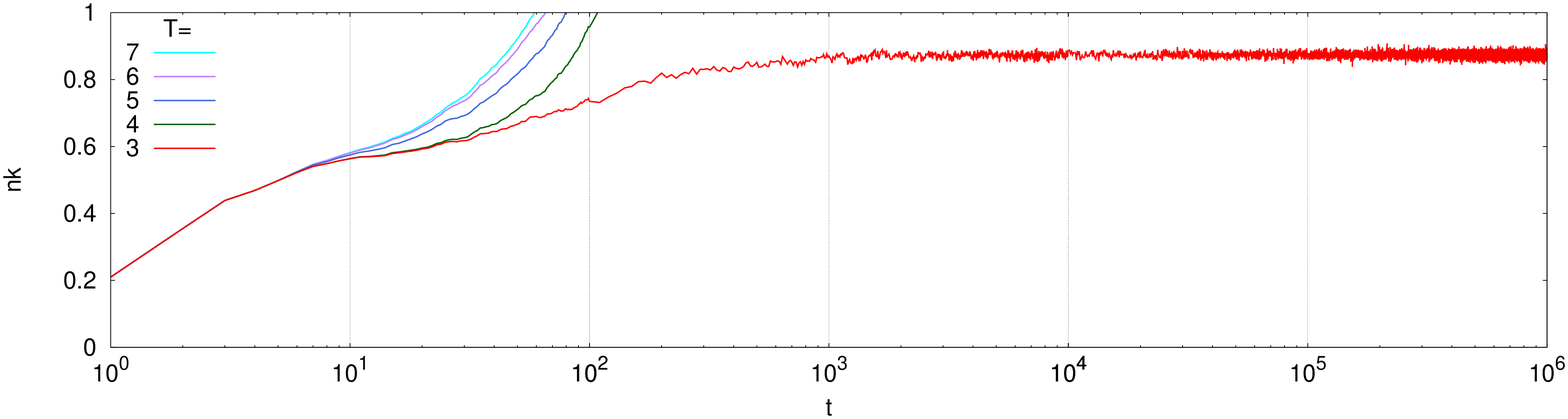}
     \caption{\label{fig-tasks-T} $M=16$, $Z=4$, $R=8$, $K=7$ and various values of $T$}
\end{subfigure}
\begin{subfigure}[t]{0.99\textwidth}
     \includegraphics[width=0.99\textwidth]{HR0_L10_M16_Z4_K7_T3_R_ntasks}
     \caption{\label{fig-tasks-R} $M=16$, $Z=4$, $T=3$, $K=7$ and various values of $R$}
\end{subfigure}
\begin{subfigure}[t]{0.99\textwidth}
     \includegraphics[width=0.99\textwidth]{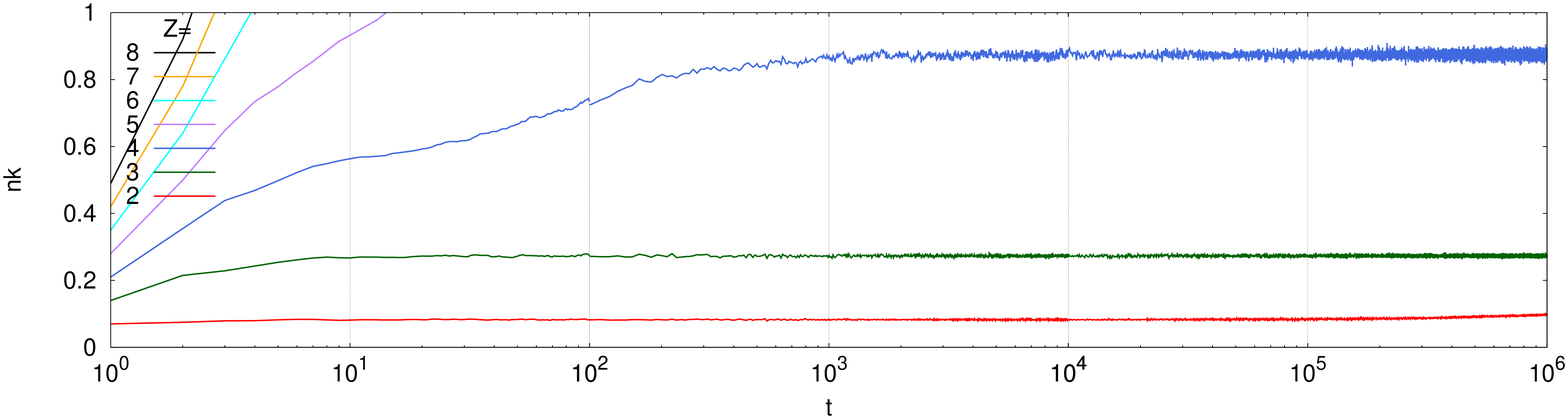}
     \caption{\label{fig-tasks-Z} $M=16$, $R=8$, $T=3$, $K=7$ and various values of $Z$}
\end{subfigure}
  \fi
\caption{\label{fig-tasks} (Color online). Time evolution of the average number $[\langle k\rangle]$ of tasks per agent obtained for \ref{eq:step6}(A) rule.
$L=10$. 
The results are averaged over $N=100$ simulations.}
\end{figure*}

The time evolution of the average fraction $[\langle\rho\rangle]$ of \change[R1]{overloaded}{loyal} agents and the average number $[\langle k\rangle]$ of tasks per agent are presented in Figs.~\ref{fig-egoists} and \ref{fig-tasks}, respectively.
The results were obtained with the rule \ref{eq:step6}(A) applied in simulation, {\em i.e.} when relaxation to \change[R1]{underloaded state}{loyalty} is possible for an {agent} who shifted out all of the tasks assigned to her/him.
The fraction of {unloyal agents} varies between $14\%<[\langle\rho\rangle]<18\%$ for $Z=4$, $R=8$, $T=3$ in `making-it' phase ($K\le 7$) [see Fig.~\ref{fig-egoists-K}].
For $K>7$ the system reaches a `jamming' phase.
Using again statistical physics terminology, we can say that on hyper-plane $(Z=4,R=8,T=3)$ the critical value of parameter $K_C=7$ separates two phases.
Similarly, the critical values of $T_C=3$, $R_C=8$ and $Z_C=4$ on hyper-planes $(Z=4,R=8,K=7)$, $(Z=4,T=3,K=7)$ and $(R=8,T=3,K=7$) may be deduced from Figs.~\ref{fig-egoists-T}, \ref{fig-egoists-R} and \ref{fig-egoists-Z}, respectively.
The same critical values of $K_C$, $T_C$, $R_C$ and $Z_C$ on mentioned above hyper-planes may be observed in Figs.~\ref{fig-tasks-K}, \ref{fig-tasks-T}, \ref{fig-tasks-R} and \ref{fig-tasks-Z} presenting time evolution of the average number $[\langle k\rangle]$ of tasks assigned to agents.
The system reaches `jammed' phase for $K>K_C$, $T>T_C$, $Z>Z_C$ and $R<R_C$.
\add[R1]{Similarly to the case of irreversible transition to the unloyalty (Sec.~}\ref{sec-irr}\add[R1]{) the role of maximal capacity parameter $M$ is quantitatively important only in this `jammed' phase.
For set of model controls parameters corresponding to `making-it' phase the changes of $M$ parameter does not influence the results of simulations (till $M$ is larger than thresholds $R$ and $T$).}

%% ----------------------------------------------------------
%% ----------------------------------------------------------

%% ----------------------------------------------------------
%% ----------------------------------------------------------

%% ##########################################################
\subsubsection{\label{sec-HRT} Relaxation to \change[R1]{underloaded state}{loyal strategy} for sufficiently low number of awaiting tasks ($k\le T$)}
%% ##########################################################

\begin{figure*}

  \if\arxiv1
\begin{subfigure}[t]{0.99\textwidth}
  \includegraphics[width=0.99\textwidth]{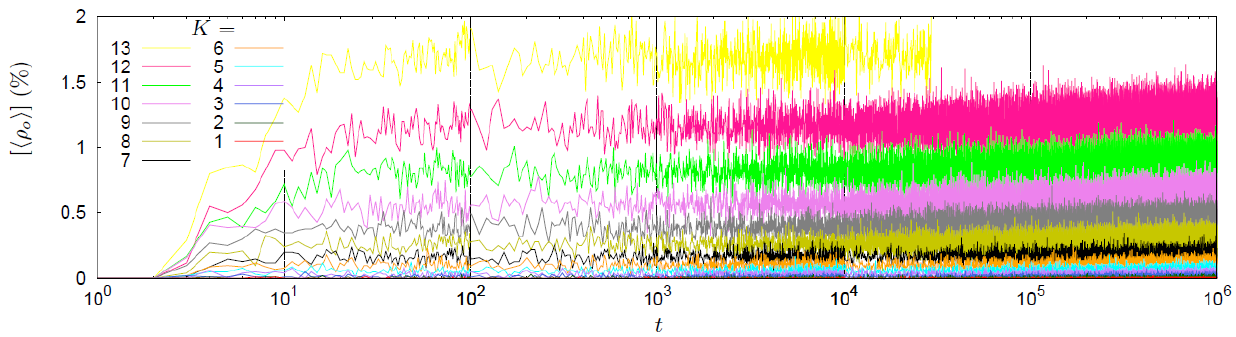}
  \caption{\label{fig-egoists-K-HRT} $M=16$, $Z=4$, $R=8$, $T=3$ and various values of $K$}
\end{subfigure}
\begin{subfigure}[t]{0.99\textwidth}
  \includegraphics[width=0.99\textwidth]{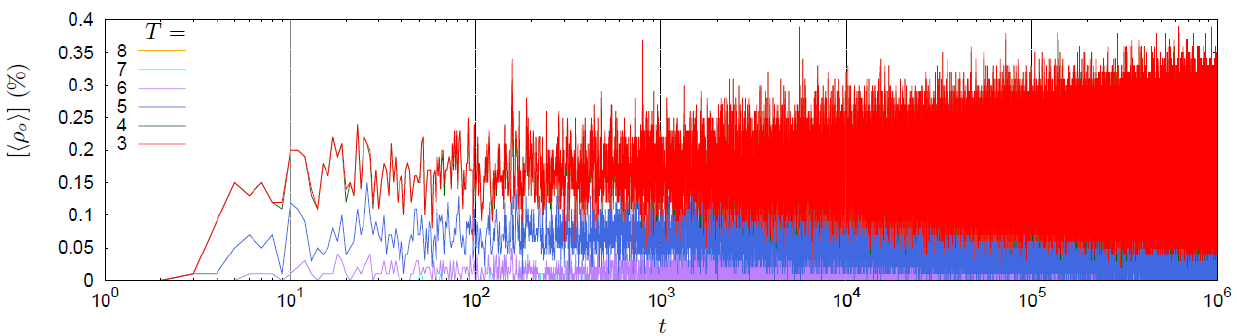}
  \caption{\label{fig-egoists-T-HRT} $M=16$, $Z=4$, $R=8$, $K=7$ and various values of $T$}
\end{subfigure}
\begin{subfigure}[t]{0.99\textwidth}
  \includegraphics[width=0.99\textwidth]{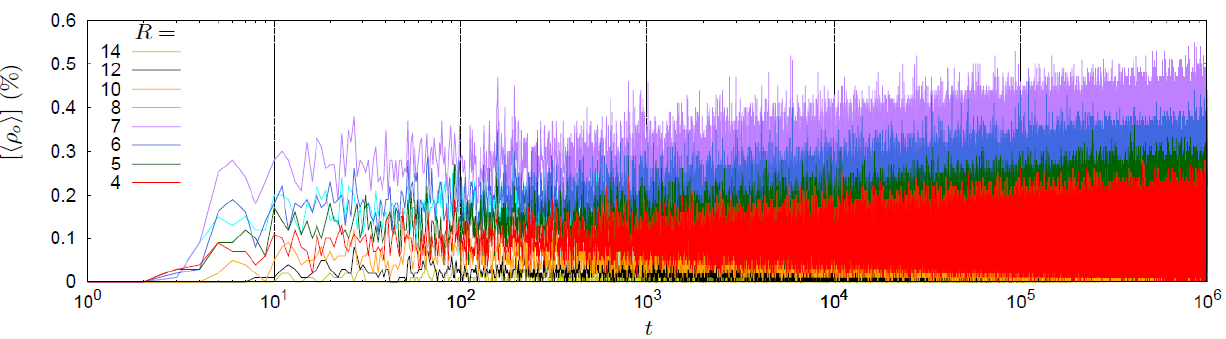}
  \caption{\label{fig-egoists-R-HRT} $M=16$, $Z=4$, $T=3$, $K=7$ and various values of $R$}
\end{subfigure}
\begin{subfigure}[t]{0.99\textwidth}
  \includegraphics[width=0.99\textwidth]{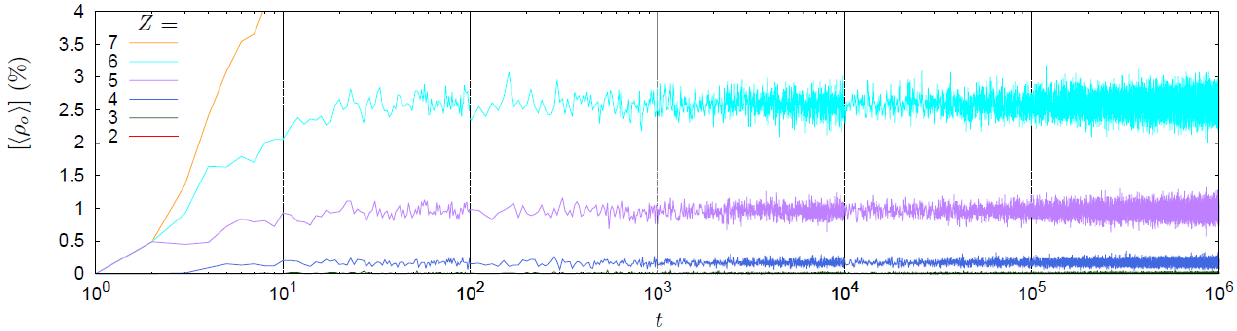}
  \caption{\label{fig-egoists-Z-HRT} $M=16$, $R=8$, $T=3$, $K=7$ and various values of $Z$}
\end{subfigure}
   \else
\psfrag{K=}{$K=$}
\psfrag{T=}{$T=$}
\psfrag{R=}{$R=$}
\psfrag{Z=}{$Z=$}
\psfrag{t}{$t$}
\psfrag{ne}[c]{$[\langle\rho\rangle]$ (\%)}
\begin{subfigure}[t]{0.99\textwidth}
  \includegraphics[width=0.99\textwidth]{HRT_L10_M16_Z4_R8_T3_K_negoists}
  \caption{\label{fig-egoists-K-HRT} $M=16$, $Z=4$, $R=8$, $T=3$ and various values of $K$}
\end{subfigure}
\begin{subfigure}[t]{0.99\textwidth}
  \includegraphics[width=0.99\textwidth]{HRT_L10_M16_Z4_R8_K7_T_negoists_new}
  \caption{\label{fig-egoists-T-HRT} $M=16$, $Z=4$, $R=8$, $K=7$ and various values of $T$}
\end{subfigure}
\begin{subfigure}[t]{0.99\textwidth}
  \includegraphics[width=0.99\textwidth]{HRT_L10_M16_Z4_K7_T3_R_negoists}
  \caption{\label{fig-egoists-R-HRT} $M=16$, $Z=4$, $T=3$, $K=7$ and various values of $R$}
\end{subfigure}
\begin{subfigure}[t]{0.99\textwidth}
  \includegraphics[width=0.99\textwidth]{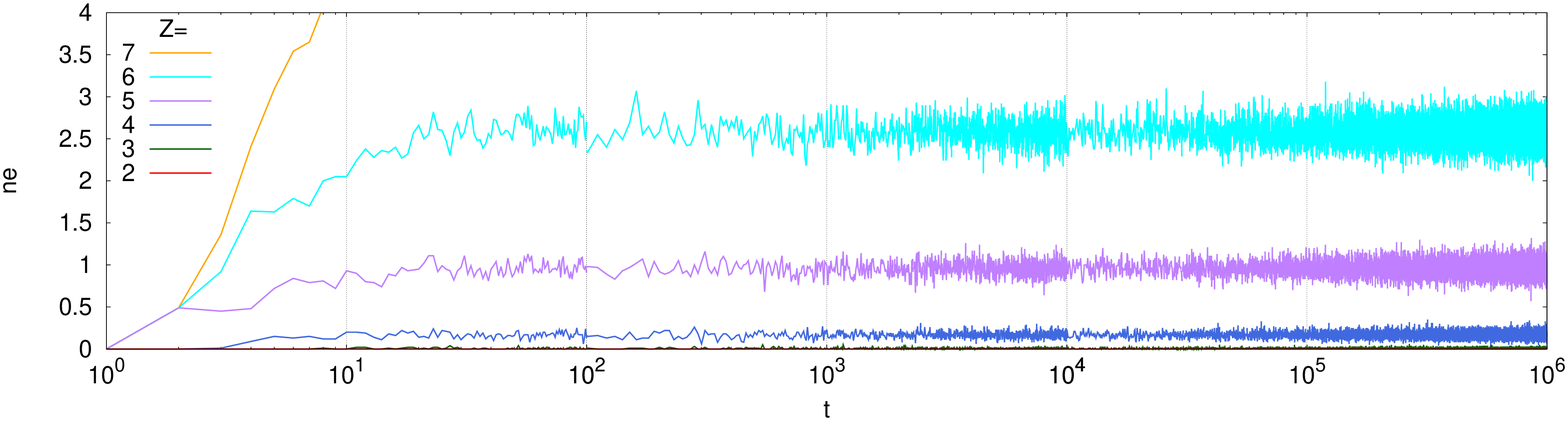}
  \caption{\label{fig-egoists-Z-HRT} $M=16$, $R=8$, $T=3$, $K=7$ and various values of $Z$}
\end{subfigure}
   \fi
\caption{\label{fig-egoists-HRT} (Color online). Time evolution of the average fraction $[\langle\rho\rangle]$ of \change[R1]{overloaded}{unloyal} agents obtained for \ref{eq:step6}(B) rule.
$L=10$.
The results are averaged over $N=100$ simulations.}
\end{figure*}

\begin{figure*}
  \if\arxiv1
\begin{subfigure}[t]{0.99\textwidth}
  {\includegraphics[width=0.99\textwidth]{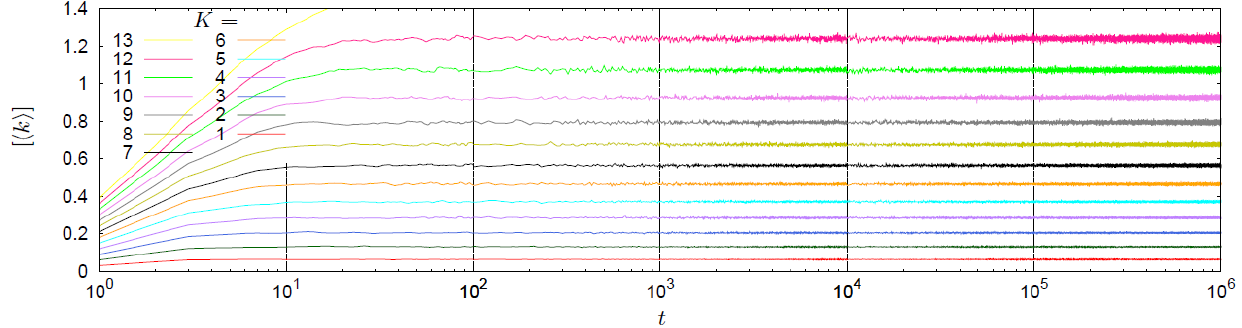}}
     \caption{$M=16$, $Z=4$, $R=8$, $T=3$ and various values of $K$\label{fig-tasks-K-HRT}}
\end{subfigure}
\begin{subfigure}[t]{0.99\textwidth}
  {\includegraphics[width=0.99\textwidth]{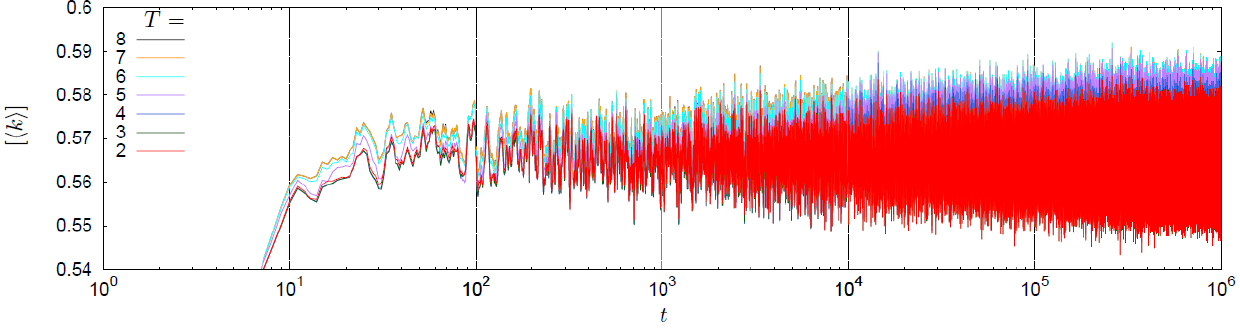}}
     \caption{$M=16$, $Z=4$, $R=8$, $K=7$ and various values of $T$\label{fig-tasks-T-HRT}}
\end{subfigure}
\begin{subfigure}[t]{0.99\textwidth}
  {\includegraphics[width=0.99\textwidth]{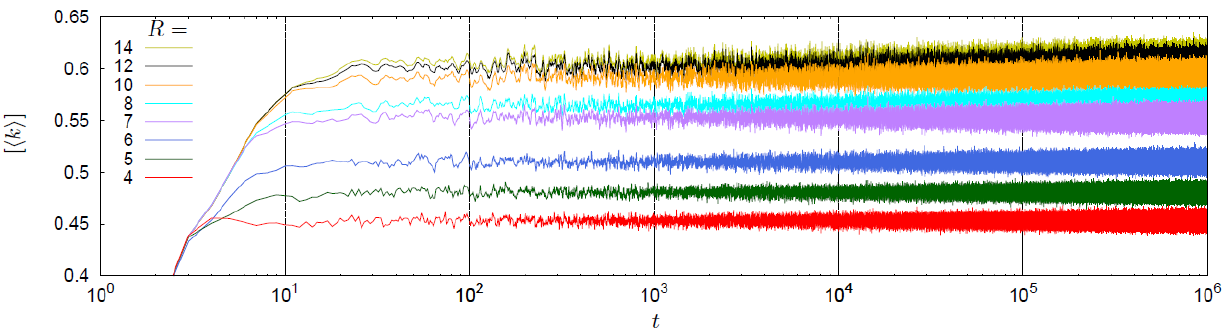}}
     \caption{$M=16$, $Z=4$, $T=3$, $K=7$ and various values of $R$\label{fig-tasks-R-HRT}}
\end{subfigure}
\begin{subfigure}[t]{0.99\textwidth}
  {\includegraphics[width=0.99\textwidth]{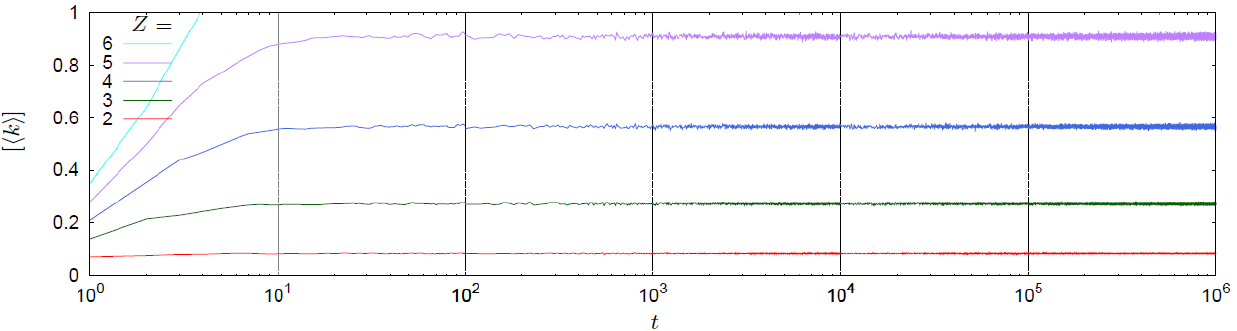}}
     \caption{$M=16$, $R=8$, $T=3$, $K=7$ and various values of $Z$\label{fig-tasks-Z-HRT}}
\end{subfigure}
   \else
\psfrag{K=}{$K=$}
\psfrag{T=}{$T=$}
\psfrag{R=}{$R=$}
\psfrag{Z=}{$Z=$}
\psfrag{t}{$t$}
\psfrag{nk}{$[\langle k\rangle]$}
\begin{subfigure}[t]{0.99\textwidth}
     \includegraphics[width=0.99\textwidth]{HRT_L10_M16_Z4_R8_T3_K_ntasks}
     \caption{$M=16$, $Z=4$, $R=8$, $T=3$ and various values of $K$\label{fig-tasks-K-HRT}}
\end{subfigure}
\begin{subfigure}[t]{0.99\textwidth}
     \includegraphics[width=0.99\textwidth]{HRT_L10_M16_Z4_R8_K7_T_ntasks_new}
     \caption{$M=16$, $Z=4$, $R=8$, $K=7$ and various values of $T$\label{fig-tasks-T-HRT}}
\end{subfigure}
\begin{subfigure}[t]{0.99\textwidth}
     \includegraphics[width=0.99\textwidth]{HRT_L10_M16_Z4_K7_T3_R_ntasks}
     \caption{$M=16$, $Z=4$, $T=3$, $K=7$ and various values of $R$\label{fig-tasks-R-HRT}}
\end{subfigure}
\begin{subfigure}[t]{0.99\textwidth}
     \includegraphics[width=0.99\textwidth]{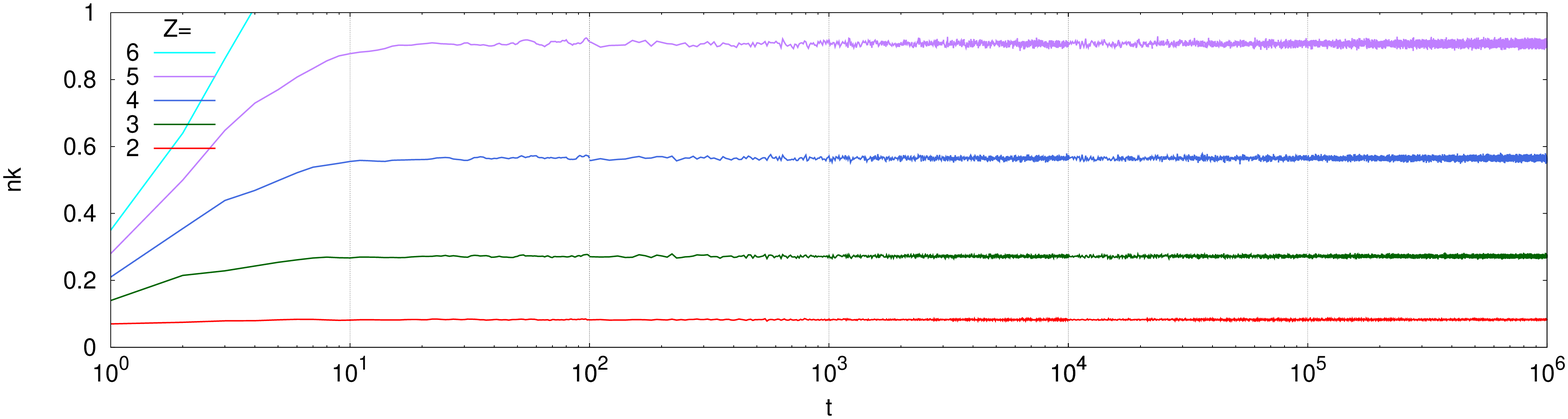}
     \caption{$M=16$, $R=8$, $T=3$, $K=7$ and various values of $Z$\label{fig-tasks-Z-HRT}}
\end{subfigure}
   \fi
\caption{\label{fig-tasks-HRT} (Color online). Time evolution of the average number $[\langle k\rangle]$ of tasks per agent obtained for \ref{eq:step6}(B) rule.
$L=10$. 
The results are averaged over $N=100$ simulations.}
\end{figure*}

When agents strategy change from \change[R1]{overloaded}{unloyal} to \change[R1]{underloaded state}{loyal one} occurs as soon as they have less than $T$ tasks the qualitative picture remains the same, {\em i.e.} we observe two regimes in tasks realization in our model {working group}.
The results presented in Figs.~\ref{fig-egoists-HRT} and \ref{fig-tasks-HRT} correspond to the step \ref{eq:step6}(B) instead of \ref{eq:step6}(A) in our algorithm.
The time evolution of the average fraction $[\langle\rho\rangle]$ of {unloyal working group members} and the average number $[\langle k\rangle]$ of tasks per agent are presented in Figs.~\ref{fig-egoists-HRT} and \ref{fig-tasks-HRT}, respectively.
The only qualitative difference appears in system response to changing $R$ and $T$ parameters: `jamming' phase occurs for small enough values of $R<R_C$ and large values of $T>T_C$ when rule \ref{eq:step6}(A) is applied and it is absent for rule \ref{eq:step6}(B).
The common difference between results of applying rule \ref{eq:step6}(A) or \ref{eq:step6}(B) is the average level of {loyalty}/{unloyalty} in {working group} and the average number of tasks awaiting for realization.
Selecting for inspection the same hyper-planes as in Sec. \ref{sec-HR0} we see definitely lower ranges of $[\langle\rho\rangle]$ and $[\langle k\rangle]$ in `making-it' phase.
These differences are presented in Tab.~\ref{tab-ranges}.
Also the change of critical parameters $K_C$, $T_C$, $R_C$ and $Z_C$ may be easily observed (see Tab.~\ref{tab-XC}).

%% ----------------------------------------------------------
\begin{table}
\caption{\label{tab-ranges} The ranges of $[\langle\rho\rangle]$ and $[\langle k\rangle]$ on $(Z=4,R=8,T=3)$, $(Z=4,R=8,K=7)$, $(Z=4,T=3,K=7)$ and $(R=8,T=3,K=7)$ hyper-planes and various variants of relaxation \change[R1]{rule} to \change[R1]{underloaded state rule}{loyalty}.
The tasks number assigned to single agent cannot exceed $M=16$.}
\begin{ruledtabular}
\begin{tabular}{cccccc}
hyper-plane & rule & \multicolumn{2}{c}{$[\langle\rho\rangle]$ (\%)} & \multicolumn{2}{c}{$[\langle k\rangle]$} \\ \hline

$(Z=4,R=8,T=3)$ & 1$-$5+6(A) & 14 & 16  & 0.4  & 0.9  \\ 
                & 1$-$5+6(B) & 0  & 0.5 & 0.5  & 2.5  \\  \hline
$(Z=4,R=8,K=7)$ & 1$-$5+6(A) & 14 & 14  & 0.9  & 0.9  \\ 
                & 1$-$5+6(B) & 0  & 0.4 & 0.55 & 0.59 \\  \hline
$(Z=4,T=3,K=7)$ & 1$-$5+6(A) & 10 & 16  & 0.75 & 0.85 \\ 
                & 1$-$5+6(B) & 0  & 0.5 & 0.45 & 0.65 \\  \hline
$(R=8,T=3,K=7)$ & 1$-$5+6(A) & 0  & 14  & 0.1  & 0.9  \\ 
                & 1$-$5+6(B) & 0  & 2.5 & 0.1  & 0.9  \\ 
\end{tabular}
\end{ruledtabular}
\end{table}
%% ----------------------------------------------------------

%% ----------------------------------------------------------
\begin{table}
\caption{\label{tab-XC} The critical values of $K_C$, $T_C$, $R_C$ and $Z_C$ on $(Z=4,R=8,T=3)$, $(Z=4,R=8,K=7)$, $(Z=4,T=3,K=7)$ and $(R=8,T=3,K=7)$ hyper-planes and various variants of relaxation \add[R1]{rules} to \change[R1]{underloaded state rule}{loyalty}.
The `jammed' phase is observed for values of $K$, $T$, $R$, $Z$ {\em larger} than $K_C$, $T_C$, $R_C$, $Z_C$, respectively.
The tasks number assigned to single agent cannot exceed $M=16$.}
\begin{ruledtabular}
\begin{tabular}{ccccc}
 & $K_C$ & $T_C$ & $R_C$ & $Z_C$ \\ \hline
rules 1$-$5+6(A): & 7  &  3 & 8\footnote{For the rule \ref{eq:step6}(A) the transition to the `jammed' state occurs for $R<R_C$.} & 4  \\
rules 1$-$5+6(B): & 12 & ---\footnote{For the rule \ref{eq:step6}(B) on the hyper-plane $(Z=4,R=8,K=7)$ only `making-it' phase is observed.} & ---\footnote{For the rule \ref{eq:step6}(B) on the hyper-plane $(Z=4,T=3,K=7)$ only `making-it' phase is observed.}  & 6  \\
\end{tabular}
\end{ruledtabular}
\end{table}
%% ----------------------------------------------------------

%% ----------------------------------------------------------
%% ----------------------------------------------------------

%% ----------------------------------------------------------
%% ----------------------------------------------------------

%% ##########################################################
\section{\label{sec-disc}Discussion and conclusions}
%% ##########################################################

In our model {working group} the crucial point for avoiding system \add[R1]{`}jamming\add[R1]{'} is allowing agents for relaxing to \change[R1]{underloaded state}{loyalty}.
In case of applying hysteretic rules [\ref{eq:step6}(A) or \ref{eq:step6}(B)] the \add[R1]{`}black scenario\add[R1]{'} of the system stacking in a \add[R1]{`}jammed\add[R1]{'} phase may be avoided when
\begin{itemize}
\item either the number of agents chosen for task realization $K$
\item or the number of assigned tasks $Z$
\item or the threshold value $R$ of assigned tasks, which force the agent to conversion from \change[R1]{underloaded}{loyalty} to \change[R1]{overloaded state}{unloyalty}
\item or the threshold value $T$ of tasks assigned to {unloyal agent}, which force him/her to task redistribution among his/her neighbors
\end{itemize}
are smartly chosen.
The term `smartly chosen' means
\begin{itemize}
\item {\em small enough} for $K$, $T$, $Z$ parameters and {\em large enough} for the parameter $R$ when rule \ref{eq:step6}(A) is applied
\item and {\em small enough} for $K$ and $Z$ parameters for the rule \ref{eq:step6}(B). 
\end{itemize}

Basically, the influence of the model parameters on the results of the simulations reflects what can be expected.
Yet, several results deserve more attention.

First one is related to the role of parameter $T$, which is the threshold; if a {unloyal} {agent} has more than $T$ tasks, he/she shifts his tasks to his/her neighbors.
When the rule \ref{eq:step6}(B) is applied the larger value of $T$, the earlier {unloyal agents} return to \change[R1]{underloaded state}{loyalty} what should result in smoother tasks realization; then it seems that the collective transition to the \change[R1]{overloaded}{unloyalty} phase should be less likely when $T$ increases.
Yet, the numerical results presented in Fig.~\ref{fig-tasks-T-HRT} indicates that the trend is slightly opposite.
On the other hand, we well understand this trend for the rule \ref{eq:step6}(A) [see Figs.~\ref{fig-egoists-T} and \ref{fig-tasks-T}].
The point is that a {unloyal} {agent} does not perform her/his tasks even if their number is less than $T$.
In this super-selfish state {agent} neither do his/her tasks nor send them to his/her neighbors.
He or she is just waiting till having again at least $T$ tasks and after receiving them above the threshold $T$ he/she redistributes them among his/her neighbors.
Please note however, once tasks are shifted, there is a chance that some of them will be performed if a neighbor is still a\remove[R1]{n} \change[R1]{underloaded}{loyal} agent.
This hypothesis may be particularly attractive if we recall clustering of \change[R1]{overloaded}{unloyal} agents presented in Fig.~\ref{egoistsxy}.
The tasks shifted by \change[R1]{overloaded}{unloyal} agents may leave the cluster of {unloyal agents} as soon as they reach its border.

The second issue is more complex: some numerical results seem to be stable, {\em i.e.} they remain constant in a certain range of at least some parameters.
An example is the percentage of \change[R1]{overloaded}{unloyal} agents, shown in Fig.~\ref{fig-egoists-K}, as dependent on the parameter $K$.
Recalling that $K$ is a measure of the amount of incoming tasks, we should expect that the amount of \change[R1]{overloaded}{unloyal} agents increases monotonously with $K$.
More exactly, the stream of incoming tasks is the product $K \cdot Z$, where $Z$ is the number of incoming tasks assigned to an {agent}. 
We could expect that the amount of \change[R1]{overloaded}{unloyal} agents should be a function of this product.
Instead, what we observe is that the amount of \change[R1]{overloaded}{unloyal} agents increases with $Z$ but remains stable with $K$, as shown in Fig.~\ref{fig-egoists-K}.
This puzzle remains to be solved.
A plausible hypothesis is that some amount of {unloyal agents} is functional as they help to transport tasks throughout the system, making the spatial distribution of tasks more homogeneous.
Four tasks can be performed by four loyal neighbors in one time step, not in four steps.
Within the range of parameters where the stability is observed, the problem of overload is maybe solved locally.
Yet, this hypothesis waits for a confirmation with dedicated numerical tools, as the local correlation functions of the task density.

The third issue is connected with ambiguous role of parameter $R$. 
It seems reasonable to expect scenario presented in Figs.~\ref{fig-egoists-R} and \ref{fig-tasks-R}, {\em i.e.} when increasing of the threshold value $R$ after which \change[R1]{underloaded}{loyal} agents become {unloyal group members} helps system in staying in `making-it' phase.
However, this natural system response to increasing $R$ is absent when rule \ref{eq:step6}(B) is applied.
Yet, for the variant \ref{eq:step6}(B), the overall amount of \change[R1]{overloaded}{unloyal} agents is much smaller.
Perhaps a weak increase of $[\langle\rho\rangle]$ and $[\langle k\rangle]$ with $R$ is related to a longer lifetime of an \change[R1]{overloaded}{unloyal} agent, who has to shift more tasks ($R-T$) to be converted to an \change[R1]{underloaded}{loyal} agent again.
Anyway, this increase seems to be a second-order effect.
On the other hand for this rule when scanning $R$ parameter we do not observe a `jammed' phase on $(Z=4,T=3,K=7)$ hyper-plane [see Figs.~\ref{fig-egoists-R-HRT}, \ref{fig-tasks-R-HRT} and Tab.~\ref{tab-XC}].

From the perspective of social simulation, the model and results reported above fall into the category of YAWOTAS (`Yet Another Way Of Thinking About Stuff'), {\em i.e.} of analogical models rather than explanatory or predictive ones \cite{13}.
Yet, this category is shared with most applications based on the technique of \change[R1]{cellular automata}{CA}.
{According to classic textbooks on social sciences} \cite{Babbie2005}, {our research can be classified as an exploration; here it deals with mathematical aspects of performance of groups of workers.
Our results suggest that the regime where their work is efficient can be stabilized by a tuning of the system parameters.}
Also we hope that the memory effect captured by hysteretic rules can be inspiring when looking solutions to other problems, even so elusive as those met in social sciences. \add[R1]{Yet we should add that our mechanism of switching to the unloyal state by passing duties from a neighbor is more specific than social contagion in general. Therefore, here we do not need to discuss the conditions of an efficient social contagion, formulated in the literature} \cite{con,hod}.

\add[R1]{The model setting and the results can be treated as a social realization of the concept of self-organized criticality} \cite{btw,bak}.
\add[R1]{When applied to a large system, these frames suggest a research on the number of `topplings', i.e. events when a given task is passed along a chain of coworkers.
In our model, tasks are irreversible and it is only their number which matters.
Yet, the unloyal strategy, when applied by the majority of agents, is equivalent to the self-organized critical state, when no task is performed; they are only passed from one agent to another. 
We note that with the rule} \ref{eq:step6}\add[R1]{(A), this collectively unloyal state is absorbing, analogously to the very idea of self-organization.
On the other hand, the scale-free distribution of the length of the above remarked chains has been discussed in Ref.} \cite{bar} \add[R1]{as a result of putting tasks with lower priority off.
These analogies promisingly link theory of organizations to current problems of statistical mechanics.}

\begin{acknowledgments}
The work was supported by the Polish Ministry of Science and Higher Education and its grants for scientific research.
\end{acknowledgments}

\end{document}